\documentclass[12pt]{article}
\usepackage{graphicx,amssymb,amsmath}
\usepackage{epsfig}
\usepackage{color,graphicx}
\usepackage{cite}
\usepackage{amssymb,amsmath}
\usepackage{multirow}
\newcommand\fverb{\setbox\pippobox=\hbox\bgroup\verb}
\newcommand\fverbdo{\egroup\medskip\noindent%
                              \fbox{\unhbox\pippobox}\ }
\newcommand\fverbit{\egroup\item[\fbox{\unhbox\pippobox}]}
\newbox\pippobox
\newcommand{\nn}{\nonumber}
\newcommand{\beq} {\begin{equation}}
\newcommand{\eeq} {\end{equation}}
\newcommand{\beqa} {\begin{eqnarray}}
\newcommand{\eeqa} {\end{eqnarray}}

\newcommand{\order}[1]{${\cal O}\left(#1 \right)$}
\newcommand{\morder}[1]{{\cal O}\left(#1 \right)}

\newcommand{\be}{\begin{equation}}
\newcommand{\ee}{\end{equation}}
\newcommand{\bea}{\begin{eqnarray}}
\newcommand{\eea}{\end{eqnarray}}

\def\om{\omega}

\begin{document}
 
\begin{flushright}
HIP-2016-40/TH
\end{flushright}

\begin{center}

\centerline{\Large {\bf Holographic sliding stripes }}

\vspace{8mm}

\renewcommand\thefootnote{\mbox{$\fnsymbol{footnote}$}}
Niko Jokela,${}^{1,2}$\footnote{niko.jokela@helsinki.fi}
Matti J\"arvinen,${}^3$\footnote{jarvinen@lpt.ens.fr} and
Matthew Lippert${}^4$\footnote{Matthew.Lippert@liu.edu}

\vspace{4mm}
${}^1${\small \sl Department of Physics} and ${}^2${\small \sl Helsinki Institute of Physics} \\
{\small \sl P.O.Box 64} \\
{\small \sl FIN-00014 University of Helsinki, Finland} 

\vspace{2mm}
${}^3${\small \sl Laboratoire de Physique Th\'eorique de l' \'Ecole Normale Sup\'erieure
\& \\ Institut de Physique Th\'eorique Philippe Meyer, \\
PSL Research University,  CNRS,  Sorbonne Universit\'es, UPMC
Univ.\,Paris 06, \\
24 rue Lhomond, 75231 Paris Cedex 05, France \\
}

\vspace{2mm}
\vskip 0.2cm
${}^4${\small \sl Department of Physics} \\
{\small \sl Long Island University} \\
{\small \sl Brooklyn, NY, United States} 

\end{center}

\vspace{8mm}

\setcounter{footnote}{0}
\renewcommand\thefootnote{\mbox{\arabic{footnote}}}

\begin{abstract}
\noindent
Holographic models provide unique laboratories to investigate non-linear physics of transport in inhomogeneous systems.  We provide a detailed account of both DC and AC conductivities in a defect CFT with spontaneous stripe order. The spatial symmetry is broken at large chemical potential and the resulting ground state is a combination of a spin and charge density wave. An infinitesimal applied electric field across the stripes will cause the stripes to slide over the underlying density of smeared impurities, a phenomenon which can be associated with the Goldstone mode for the spontaneously broken translation symmetry.   We show that the presence of a spatially modulated background magnetization current thwarts the expression of some DC conductivities in terms of horizon data.
\end{abstract}

\newpage
\tableofcontents


\newpage
\section{Introduction}

Many condensed matter systems exhibit ordered phases in which the microscopic translation or lattice symmetry is spontaneously broken.  Such states are particularly interesting because they often feature highly unusual properties, especially transport properties, which could potentially lead to novel practical applications.

Striped phases arise in a variety of contexts.  As early as 1930, Peierls predicted that quasi-one-dimensional metals could become unstable at low-temperature, leading to charge modulated phases \cite{Peierls}.  Such charge density wave (CDW) states are characterized by the condensation of an electron-hole bilinear.  CDW states can have strikingly nonlinear DC conductivity.  Lattice charges or localized impurities tend to pin the CDW in place, but above a threshold electric field, the CDW begins to slide, carrying a collective, often highly non-uniform electric current. CDW phases have been observed in many materials, although often the pinned effect of impurities is too strong for collective transport.\footnote{For a review of CDW physics, see \cite{Gruner1}.}

Spatially modulated phases also appear, for example, in extremely pure two-dimensional electron liquids in weak magnetic fields where several Landau levels are filled.  For large non-integer filling fractions, electron-electron interactions drive the system toward the creation of CDW states \cite{Fogler}.

A spin density wave (SDW) is a one-dimensional, antiferromagnetic symmetry-breaking ground state found in some metals, notably chromium \cite{Cr SDW}, and also in certain organic linear-chain conductors.  Like CDW states, a SDW features a coherent electron-hole bilinear, but with opposite spins.  A SDW can alternately be viewed as two opposite-spin, out-of-phase CDWs.  SDW phases exhibit both collective charge and spin excitations.\footnote{For a review of SDW states, see \cite{Gruner2}.}

Typically, striped phases occur in weakly coupled systems amenable to a perturbative description in terms of particle-like excitations.  In other cases, however, striped phases have been observed in strongly coupled materials.  For example, the poorly understood pseudogap regime of underdoped cuprate superconductors features a variety of striped states.\footnote{For a review of striped phases in cuprate superconductors, see \cite{Vojta,Kivelson:2003zz}.}

Holographic systems with broken translation invariance, and striped phases in particular, have been a focus of much attention in recent years. 
If momentum is conserved as a result of translation invariance, currents have no way to disperse, resulting in strictly infinite DC conductivity, which is clearly unphysical. A mechanism for dissipation is needed to obtain realistic transport properties.
One motivation for these holographic studies is indeed to provide a mechanism for momentum dissipation. 
	
The most common approach has been to break translation symmetry explicitly. Introducing a spatially modulated source, such as a periodic chemical potential, can be interpreted as modeling an underlying lattice \cite{Aperis:2010cd, Hutasoit:2011rd, Ganguli:2012up, Liu:2012tr,Hutasoit:2012ib, Horowitz:2012gs, Ganguli:2013oya, Erdmenger:2013zaa, Ling:2013aya, Ling:2013nxa} or disorder \cite{Arean:2013mta,  Arean:2014oaa}, while a single localized source represents a defect \cite{HoyosBadajoz:2010ac,Ammon:2012dd,Araujo:2015hna}.  Linearly varying scalar fields cause the weakest breaking, yielding a homogeneous stress tensor and allowing a more tractable analysis \cite{Vegh:2013sk, Davison:2013jba,Blake:2013bqa,Blake:2013owa, Andrade:2013gsa, Amoretti:2014zha, Gouteraux:2014hca,Taylor:2014tka, Donos:2014oha}.  When symmetry-breaking sources are included, holographic models can produce Drude-like conductivities \cite{Horowitz:2012ky,Donos:2014yya}.  However, when the inhomogeneities are added by hand, much of the novel physics associated with CDWs and other striped phases is absent.  

Striped phases are interesting primarily because the spatial ordering is generated spontaneously by competing interactions.  In several holographic models \cite{Nakamura:2009tf, Ooguri:2010kt, Donos:2011bh, Donos:2011qt, Bu:2012mq, Rozali:2013ama, Domokos:2007kt, Ooguri:2010xs, Chuang:2010ku, Bayona:2011ab, BallonBayona:2012wx, Bergman:2011rf, Jokela:2012vn, Jokela:2012se}, the homogeneous solutions suffer from a modulated instability.  In most cases, the instability is essentially that of Maxwell-Chern-Simons theory with a constant electric field \cite{Nakamura:2009tf}.\footnote{Different types of spontaneously modulated instabilities  preserving both parity and time reversal invariance were found in \cite{Donos:2013gda,Gouteraux:2016arz}.}  The resulting striped solutions have been found in some cases \cite{Ooguri:2010kt, Ooguri:2010xs, Bayona:2011ab, Donos:2012gg, BallonBayona:2012wx, Bu:2012mq, Rozali:2012es, Donos:2012yu,Donos:2013wia, Withers:2013loa, Withers:2013kva, Rozali:2013ama, Ling:2014saa, Withers:2014sja,Jokela:2014dba,Donos:2015eew,Donos:2016hsd,Amoretti:2016bxs}, but the novel transport properties of these states remains to be thoroughly analyzed.

With that goal in mind, we will focus on the D3-D7' probe-brane model \cite{Bergman:2010gm, Jokela:2010nu, Bergman:2011rf, Jokela:2012vn, Jokela:2013hta, Jokela:2014wsa}, a well-studied example for which the spontaneously striped solution was explicitly computed in \cite{Jokela:2014dba}.  Above a critical charge density, there is a second-order phase transition from the gapless, homogeneous state to a spin and charge density wave state with modulated persistent currents parallel to the stripes.  A background magnetic field stabilizes the homogeneous state, increasing the critical charge density.  In addition, at nonzero magnetic field the phase transition becomes first order.

The DC charge conductivity of the D3-D7' model in the homogeneous phase was computed in \cite{Bergman:2010gm} using the Karch-O'Bannon method \cite{KOB}.  Even though momentum is conserved, probe brane models have finite DC conductivity. The D3-branes act like impurities smeared within the worldvolume of the D7-branes, and, because the D7-branes are treated as probes in the D3-brane background, the $\mathcal N=4$ SYM sector serves as an effective momentum sink for the charged matter sector.

In the gapless phase, corresponding holographically to a black hole embedding of the D7-brane, the charge is distributed into two fluids:  fractionalized charges at the black hole horizon and cohesive charges smeared radially along the D7-brane. The Hall conductivity is the sum of two contributions, one from each fluid.  The longitudinal conductivity, however, involves only the charges at the horizon and a contribution from pair production.  The horizon charges interact with the $\mathcal N=4$ SYM sector and are dissipative, while the induced charges are 
dissipationless.

In this paper, we study the AC/DC conductivities of the spatially modulated striped phase possessing both CDW and SDW. We choose parameter values that do not explicitly break parity nor time invariance of the gapless state. In particular, we consider a special case of the background constructed in~\cite{Jokela:2014dba} with zero magnetic field.
Under these conditions the corresponding homogeneous state has been argued to resemble graphene \cite{Davis:2011gi,Omid:2012vy,Semenoff:2012xu}. The possibility of obtaining CDW and SDW phases within graphene-like systems under strong interactions was proposed in \cite{Araki:2012gs}. Our results could therefore find application in such a context.

Our most striking result is that an applied electric field causes the stripes to move.  Without any localized impurities or underlying lattice explicitly breaking translation invariance, the stripes are not pinned to any particular location. We find that the stripes have a sliding zero mode, which is precisely the Goldstone mode for the spontaneously broken translation symmetry. The stripes are not pinned to the smeared impurities represented by $\mathcal N=4$ SYM sector but slide with a velocity proportional to the applied electric field.

The striped phase spontaneously breaks parity in a modulated way, and we find the Hall conductivity in the striped phase reflects this breaking.  An electric field applied across the stripes generates a modulated Hall current in the parallel direction.  In particular, there is a pole in the DC Hall conductivity related to the persistent modulated currents of the background striped solution.  However, because the stripes are parity symmetric on average, the Hall conductivity vanishes when averaged over the modulated direction. On the other hand, an electric field applied parallel to the stripes does not produce a Hall current across them at all.

This paper is organized as follows.  In Sec.~\ref{sec:background}, we review the D3-D7' model, including the homogeneous and striped solutions.   Then, in Sec.~\ref{sec:DC conductivity}, we  analytically compute the DC conductivities for the stripe phase in terms of horizon data.  We turn next in Sec.~\ref{sec:AC conductivity} to a numerical computation of the AC conductivity.  Sec.~\ref{sec:Discussion} contains a discussion of the goal of finding a fully nonlinear sliding stripe solution, and we conclude with a brief discussion of some open problems in  Sec.~\ref{sec:Conclusion}. Details of the analysis of the DC conductivities are given in App.~\ref{app:DCcond}, and the numerical solutions of the fluctuation equations and their approximate $x \leftrightarrow y$ symmetries are discussed in App.~\ref{app:fluct}.


\section{Background}
\label{sec:background}

The D3-D7' model consists of a probe D7-brane in a D3-brane background such that the intersection is a (2+1)-dimensional defect which completely breaks supersymmetry and whose low-energy excitations are purely fermionic.\footnote{This is a member of a family of well studied \#ND=6 brane intersection models~\cite{Sakai:2005yt,Sakai:2004cn,Jokela:2011eb,Jokela:2011sw,Jokela:2015aha}.}  This system was originally designed as a holographic model of the quantum Hall effect because it features a Minkowski embedding at nonzero charge density, yielding a gapped, quantum Hall phase \cite{Bergman:2010gm, Jokela:2010nu}.  In this paper, we focus instead on the black hole embedding, which is dual to a gapless, conducting quantum fluid \cite{Bergman:2010gm, Bergman:2011rf, Jokela:2012vn, Jokela:2014dba, Davis:2011gi, Omid:2012vy}.

The ten-dimensional black D3-brane metric reads:
\be
 ds_{10}^2 = \frac{r^2}{L_{AdS}^2}\left(-h(r)dt^2+dx^2+dy^2+dz^2\right)+\frac{L_{AdS}^2}{r^2}\left(\frac{dr^2}{h(r)}+r^2d\Omega_5^2\right) \ ,
\ee
where the thermal factor is $h=1-\left(\frac{r_T}{r}\right)^4$, which is related to the temperature as $r_T=\pi T$. We write the metric on the internal sphere as an $S^2\times S^2$ fibered over an interval:
\be
 d\Omega_5^2 = d\psi^2+\cos^2\psi\left(d\theta^2+\sin^2\theta \ d\phi^2\right)+\sin^2\psi\left(d\alpha^2+\sin^2\alpha \ d\beta^2\right) 
\ee
with the following ranges for angles: $\psi\in [0,\pi/2]$; $\theta,\alpha\in [0,\pi]$; and $\phi,\beta\in [0,2\pi]$. Recall also that the Ramond-Ramond four-form is $C^{(4)}_{txyz}=-r^4/L_{AdS}^4$, and the curvature radius
is related to the 't Hooft coupling as $L_{AdS}^2=\sqrt{4\pi g_s N}\alpha' = \sqrt{\lambda}\alpha'$. We will work in units where $L_{AdS}=1$.

The probe D7-brane is embedded so that it spans the $t,x,y$ Minkowski directions, is extended in the holographic radial direction $r$, and wraps both of the two-spheres.  The background solutions are specified by the embedding functions $z(t,x,y,r)$ and $\psi(t,x,y,r)$. 
We  need the following internal components of the gauge field in order to stabilize the embedding against slipping off the internal $S^5$:
\bea
 2\pi\alpha' F_{\theta\phi} & = & \frac{f_1}{2}\sin\theta \\
 2\pi\alpha' F_{\alpha\beta} & = & \frac{f_2}{2}\sin\alpha \ .
\eea
The quantized constants $f_1$ and $f_2$ are proportional to the number of dissolved D5-brane flux on the internal two-spheres. For $f_1$ and $f_2$ within a particular range, the tachyonic mode is lifted above the BF bound \cite{Bergman:2010gm}. 

The D7-brane action consists of a Dirac-Born-Infeld term and a Chern-Simons term:
\be 
 S  =  -T_7 \int d^8x\, e^{-\Phi} \sqrt{-\mbox{det}(g_{\mu\nu}+ 2\pi\alpha' F_{\mu\nu})} -\frac{(2\pi\alpha')^2T_7}{2} \int P[C_4]\wedge F \wedge F \ . \label{totalact}
\ee
This action enjoys discrete $\mathcal{P}$ and $\mathcal{C}$ symmetries, given as~\cite{Omid:2012vy}
\begin{align}
 \mathcal{P}: \quad (t,x,y) &\mapsto (t,-x,y) \ , & \quad  (a_t,a_x,a_y,a_r) &\mapsto (a_t,-a_x,a_y,a_r)\ ,& \nn\\
  \psi &\mapsto \frac{\pi}{2} - \psi \ ,&  (f_{1},f_2) &\mapsto (f_2,f_1) \ ;&  \\
 \mathcal{C}: \quad (t,x,y) &\mapsto (t,x,y) \ , & \quad (a_t,a_x,a_y,a_r) &\mapsto (-a_t,-a_x,-a_y,-a_r)\ ,& \nn\\
  \psi &\mapsto \psi \ ,&  (f_{1},f_2) &\mapsto (f_1,f_2) \ ;&
\end{align}
and with $z$ remaining invariant under both transformations. Naturally, the action is also symmetric under arbitrary rotations of the  $(x,y)$-plane, which act on the vector $(a_x,a_y)$ as well.

Here we will only consider the case $f_1=f_2=1/\sqrt{2}$, for which a striped instability was found in~\cite{Bergman:2011rf}, and the end point of this instability was constructed numerically in~\cite{Jokela:2014dba}. For generic embeddings the action is complicated, but for (possibly modulated) backgrounds depending only on $x$ and $r$ it can be written in a relatively compact form:
\bea
 S & = & -{\cal{N}}_T\int d\hat x du \ u^{-2}\bigg[\sqrt G\sqrt{u^{-4} \hat A+\hat A_{x}+\hat A_{xu}}-\frac{1}{2}u^{-2} \partial_u\hat z \nonumber\\
  & & +2 c(\psi)u^2( \partial_u\hat a_t\partial_{\hat x}\hat a_{y}-\partial_{\hat x}\hat a_t\partial_u\hat a_{y})\bigg] \ ,\label{eq:uaction}
\eea
where
\bea
 G & = & \frac{1}{4}\left(1+8\cos^4\psi\right)\left(1+8\sin^4\psi\right) \\ 
 \hat A & = & 1+h u^2 \partial_u\psi^2+h \partial_u\hat z^2-u^4 \partial_u\hat a_t^2+h u^4\partial_u\hat a_y^2 \\
\label{Axexp} 
\hat A_{x} & = & -\frac{1}{h}\partial_{\hat x}\hat a_t^2+\partial_{\hat x}\hat a_{y}^2+u^{-4}\partial_{\hat x}\hat z^2+u^{-2}\partial_{\hat x}\psi^2 \\
 \hat A_{xu} & = & -u^4\partial_{\hat x}\hat a^2_t\partial_u\hat a_y^2-\partial_u\hat z^2\partial_{\hat x}\hat a^2_t-u^2\partial_u\psi^2\partial_{\hat x}\hat a^2_t-u^{4}\partial_u\hat a_t^2\left(u^{-4}\partial_{\hat x}\hat z^2+u^{-2}\partial_{\hat x}\psi^2+\partial_{\hat x} \hat a_{y}^2\right) \nonumber \\
 & & +h \partial_u\hat z^2\partial_{\hat x}\hat a_{y}^2+h u^2\partial_u\psi^2\partial_{\hat x}\hat a_{y}^2+h u^{-2}\partial_u\psi^2\partial_{\hat x}\hat z^2+h\partial_u\hat a_y^2\partial_{\hat x} z^2+h u^{-2}\partial_u\hat z^2\partial_{\hat x}\psi^2+ \nonumber \\
 & & h u^2 \partial_u\hat a_y^2\partial_{\hat x}\psi^2 -2h u^{-2}\partial_u\hat z \partial_u\psi\partial_{\hat x}\hat z\partial_{\hat x}\psi-2 h\partial_u\hat a_y\partial_{\hat x} \hat a_{y}(\partial_u\hat z\partial_{\hat x}\hat z+u^2\partial_u\psi\partial_{\hat x}\psi) \nonumber \\
 & & +2u^2\partial_u\hat a_t \partial_{\hat x}\hat a_t\left(u^2\partial_u\hat a_{y}\partial_{\hat x}\hat a_{y}+u^{-2}\partial_u\hat z\partial_{\hat x}\hat z+\partial_u\psi\partial_{\hat x}\psi \right) \ ,
\eea
and
\be
 c(\psi) = \psi-\frac{1}{4}\sin(4\psi)- \frac{\pi}{4} \ . 
\ee
The overall factor is ${\cal{N}}_T=4\pi^2 r_T^2 T_8 V_{1,1}$. We have introduced a new radial coordinate $u$:
\be
 u = \frac{r_T}{r} \ ,
\ee
which casts the location of the horizon to $u=1$ and the $AdS$ boundary to $u=0$. 
We have also stripped off the explicit dependence of $r_T$ factors by introducing a notation with hats as follows:
\be
 \hat x^\mu = \frac{x^\mu}{r_T}\quad ,\quad \hat z=\frac{z}{r_T}\quad ,\quad \hat a_\mu = \frac{a_\mu}{r_T} \ .
\ee
This means that all of the parameters are written in units of the horizon radius $r_T$. However, to streamline the notation we will drop the hats from now on. Notice also that as the gauge potential $a_x$ is decoupled, it was consistent to set it to zero. Further, we chose the radial gauge $a_u =0 $.

This action \eqref{eq:uaction} admits solutions which spontaneously break translation invariance through the formation of stripes, such that the boundary conditions (sources) are independent of $x$ but subleading terms at the boundary and the full bulk solutions depend on it. In general, such solutions depend on the values of the sources for the embedding and the gauge field: the (rescaled) chemical potential $\mu = a_t(x,u=0)$, the quark mass $m = \partial_u \psi (x,0)$, and the magnetic field $b$. For simplicity, here we will only consider solutions with massless quarks and no background magnetic field: $m=0=b$. 
These solutions are black hole embeddings, as the D7-brane probe reaches the horizon and which corresponds to a gapless metallic state.  The embedding coordinates and D7-brane gauge fields are functions of the spacetime coordinate $x$ as well as the radial direction $u$: $\psi=\psi(x,u),  z = z(x, u), a_t = a_t(x, u)$, and $a_y =  a_y(x, u)$. This striped solution was found and analyzed in~\cite{Jokela:2014dba}.  The purpose of this paper is to analyze the transport properties of this modulated background embedding.

All fields are nontrivially coupled with each other in the striped background solution and have nonzero modulation in the $x$-direction. Among the various worldvolume fields, the amplitude of the modulation varies significantly. The leading modulation appears in the embedding $\psi$ and the gauge field $a_y$, implying a modulated bilinear condensate, electric current in $y$-direction, and magnetization. This leading effect was therefore identified as the SDW in~\cite{Jokela:2014dba}. The temporal component of the gauge field, and the charge density are also modulated, but the amplitude of modulation is suppressed by approximately two orders of magnitude. This subleading modulation was identified as the CDW.

The striped solutions conserve a subset of the underlying symmetries of the action~\eqref{totalact}. Charge conjugation symmetry $\mathcal{C}$ is broken due to finite chemical potential, but the solutions conserve parity symmetry $\mathcal{P}$ under reflection about a fixed axis (in the $x$-direction) when the magnetic field equals zero. Moreover, the solutions are symmetric in the rotation of the $(x,y)$-plane by 180 degrees, where the overall sign of $(a_x,a_y)$ is also flipped. As in \cite{Jokela:2014dba}, we have chosen the $x$-coordinate such that the rotation symmetry is realized as rotation around the origin on the $(x,y)$-plane, whereas the parity symmetry is realized as a reflection about the axis $x=L/4$ where $L$ is the periodicity of the solutions in the $x$-direction. Explicitly, the symmetries of the background are therefore
\begin{align}
\label{symmbg180rot}
 (\psi,z,a_t,a_y) &\mapsto (\psi,z,a_t,-a_y) \ ,&\quad x &\mapsto -x \ ;& \\
\label{symmbgparity}
 (\psi,z,a_t,a_y) &\mapsto (\pi/2-\psi,z,a_t,a_y) \ ,&\quad x &\mapsto L/2-x \ , & 
\end{align}
and the latter of these symmetries (parity) would be broken at finite magnetic field. The symmetries of the fluctuations about this background will be discussed in Sec.~\ref{sec:AC conductivity}.

We will now discuss the slightly nontrivial division of charge into two components. The physical charge density is defined as
\be
D(x) = - \left.\frac{\delta S}{\delta \partial_u a_t} \right|_{{\rm{boundary}}} \ ,
\ee
and as usual, found to be proportional to the rescaled charge density\footnote{Notice that our normalization for $d(x)$ differs from that of~\cite{Jokela:2014dba} by a factor of $\sqrt{2}$.}
\be
d(x) = - \partial_u a_t |_{{\rm{boundary}}} \ ,
\ee
which we will simply call the charge density below.
In order to study the components of the charge, and for later use in the analysis of the conductivities, we need to
derive an expression for the averaged charge density in terms of the horizon data.  We will use the following notation for averaged quantities:
\be\label{eq:average}
 \langle \cdots \rangle = \frac{1}{L}\int_0^L dx (\cdots) \ .
\ee
First we note that the action depends on $a_t$ only through its derivatives. As the striped solutions only depend on $u$ and $x$, the relevant derivatives are $\partial_u a_t$ and $\partial_x a_t$. Integrating the $a_t$ equation of motion over the period of the stripes, we obtain
\be \label{conservedquantity}
 \partial_u \left\langle\frac{\delta S}{\delta ( \partial_u a_t(x,u) )} \right\rangle = 0 
\ee
since the other term averages to zero. 
At the horizon, we expand the background solution as
\begin{align}
 \psi(x,u) &= \psi_0(x) + \morder{1-u}\ ,& \qquad  z(x,u) &= z_0(x) + \morder{1-u}\ ,& \\
  a_y(x,u) &= a_{y,0}(x) + \morder{1-u}\ ,&\qquad a_t(x,u) &= a_{t,0}(x)(u-1) +\morder{(1-u)^2} \ .
\end{align}
Evaluating the conserved quantity in (\ref{conservedquantity}) in the UV and in the IR gives the (total) averaged charge density in terms of horizon data: 
\be \label{chargeidentity}
\langle d \rangle 
= \sqrt{2} \langle c(\psi_0) a_{y,0}' \rangle -  \langle a_{t,0} \hat \sigma \left(1+a_{y,0}'(x)^2+\psi_0'(x)^2+z_0'(x)^2\right)\rangle \ , 
\ee
where the prime denotes the derivative with respect to $x$ and
\be 
 \hat\sigma(x) = \frac{ \sqrt{\left(1+8 \sin ^4 \psi_0(x)\right) \left(1+8 \cos ^4 \psi_0(x)\right) }}{2\sqrt{2\left(1- a_{t,0}(x)^2\right) \left(1+a_{y,0}'(x)^2+\psi_0'(x)^2+z_0'(x)^2\right) }} \ .  
 \label{localcond1}
 \ee
Notice that for homogeneous backgrounds, this expression reduces to the standard formula for the DC conductivity~\cite{Bergman:2010gm,Hutchinson:2014lda}. In Sec.~\ref{sec:DC conductivity} we will demonstrate that $\hat \sigma$ is connected to the DC conductivities and discuss the analogy to classical circuits.

It is tempting to identify the first term on the right hand side of~\eqref{chargeidentity} as the induced charge density and the second term as the ordinary charge density: The first term  arises due to the contribution of the Chern-Simons term in~\eqref{conservedquantity}, as is expected for induced charge. The second term is proportional to the derivative of $a_t$ at the horizon suggesting that it counts the charge cloaked by the horizon. We expect that the division to the induced and ordinary charge will become more clear after turning on a nonzero magnetic field. Notice that the first term is nonzero because of the SDW component of the stripes, which corresponds to the modulation of $\psi$ and $a_y$. For the homogeneous solution we have that both $c(\psi_0) = 0 = a_{y,0}'$.

\begin{figure}[!ht]
\center
 \includegraphics[width=0.30\textwidth]{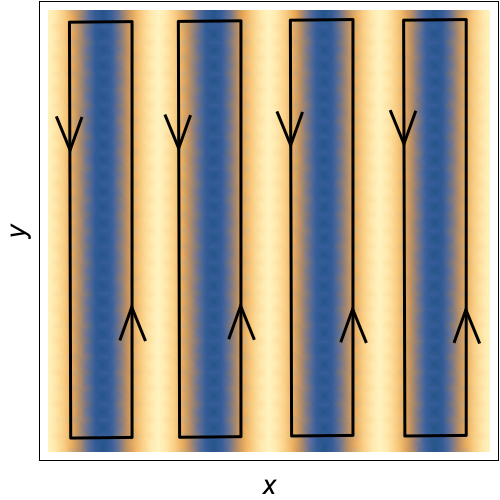}
  \caption{A sketch of the circulating currents present in the striped solution. To mimic a realistic system, we have implemented fake edges in the $y$-direction such that the currents circulate around each stripe.} 
  \label{fig:stripy}
\end{figure}

Before we investigate the linear response to external, and small, electric field applied on the stripes, it is instructive to note the following property of the background. Recall that the striped solution is such that $\partial_u a_y(x,u\to 0)\equiv J_y(x) \ne 0$. This has the interpretation of a local anomalous modulated Hall current, as it is non-vanishing without any applied electric field. Alternatively, the current $J_y(x)$ can also be called the magnetization current. However, since $\langle J_y(x)\rangle=0$, and $J_x=0$, there is no net flow of charge carriers across the sample. Instead, we can easily posit the following interpretation of the situation. There are circulating currents around each stripe, connected at spatial $y$-infinities. We have sketched this in Fig.~\ref{fig:stripy}, where we have imposed fake edges at $y$-boundaries.  Notice that since there is a non-vanishing current in the background without an external electric field, one should expect to find an infinite local DC Hall conductivity 
$|\sigma_{yx}(x)|\to \infty$ (but with $\langle \sigma_{yx}(x)\rangle=0$). This is indeed the case, and we will discuss this in more detail in the following sections.

\section{DC conductivities}
\label{sec:DC conductivity}

In this section we investigate the direct current electrical conductivities of the striped solution introduced in the previous section. DC conductivities of holographic systems have been discussed extensively in the literature. For homogeneous solutions in probe brane systems they can be computed in terms of background horizon data even at finite electric fields~\cite{KOB}. Recently a generic formalism has been developed that expresses the averaged thermoelectric DC conductivities for inhomogeneous backgrounds in the context of Einstein-Maxwell theories (see~\cite{Donos:2015gia,Banks:2015wha,Donos:2015bxe}). We will discuss below in detail our case: DC conductivities for probe branes in a background where translation symmetry is spontaneously broken in one direction. DC conductivities in a rather similar configuration, an interface which explicitly breaks translation invariance in one direction, had been recently discussed in~\cite{Araujo:2015hna}. However, when the translation invariance is spontaneously broken, the Goldstone mode corresponding to the ``sliding'' background also needs to be taken into account.

It is useful to sketch first what can be computed in our case and why. Expressing the DC conductivities in terms of horizon data requires that there is a conserved ``current'' in the bulk which links the electric currents on the boundary to the values of the background on the horizon. We have chosen coordinates such that the background depends on $x$ but not on $y$, and the same will hold for the relevant fluctuations also. Because the Lagrangian depends on the gauge fields $a_u$ and $a_x$ only through the derivatives $\partial_x a_u$ and $\partial_u a_x$, respectively, their equations immediately imply that 
\be
 \frac{\partial \mathcal{L}}{\partial \left(\partial_x a_u\right)} = - \frac{\partial \mathcal{L}}{\partial \left(\partial_u a_x\right)}
\ee
is independent of $x$ and $u$. This holds both for the background fields and their fluctuations, and in the latter case gives the sought conserved quantity for the electric current in the $x$-direction. This allows us to express the conductivities $\sigma_{xx}$ and $\sigma_{xy}$ in terms of the horizon data independently for all values of $x$. We will show below that this holds even when the possible movement of the background (or the Goldstone mode) is taken into account.

The same arguments do not apply for the current in the $y$-direction. This is the case because the Lagrangian depends on $a_y$ both through $\partial_x a_y$ and $\partial_u a_y$, so that the fluctuation equation for $a_y$ contains two terms and cannot be integrated. The term arising from the derivative with respect to  $\partial_x a_y$ can, however, be eliminated by averaging the fluctuation equation over a period in the $x$-direction, implying that an averaged conserved current in the bulk exists. Consequently, as we will show in detail below, we can compute the averaged conductivities $\langle \sigma_{yy}\rangle$ and  $\langle \sigma_{yx}\rangle$ in terms of horizon data.

\subsection{Electric field perpendicular to stripes}

We first study the current in the $x$-direction with an electric field turned on only in this direction, so that we can analyze $\sigma_{xx}$. 
We will discuss the other components of the DC conductivity matrix in Sec.~\ref{sec:jyandEy}.
Turning on the electric field perpendicular to the stripes, we expect that the stripes begin to slide, and therefore the Goldstone modes contributes to $\sigma_{xx}$.

In order to compute $\sigma_{xx}$, we study time-dependent fluctuations around the striped solution having the form $\delta f (t,x,u)$ where $f = \psi$, $z$, $a_t$, $a_y$, $a_x$. We have fixed the gauge such that the fluctuation of $a_u$ vanishes. 
Because the translation invariance is spontaneously broken in the striped phase, there is indeed a Goldstone mode which is given by the $x$-derivative of the background solution: the (time-independent) Ansatz $\delta f(t,x,u) = \kappa \partial_x f(x,u)$. 
The Goldstone mode has this form because for any background solution $f(x,u)$ with $x$-independent sources, $f(x+\kappa,u)$ is a solution as well. The variation, which is obtained by expanding at small $\kappa$ and nonvanishing for inhomogeneous solutions, is identified as the Goldstone mode, and it automatically solves the fluctuation equations.

As we have argued above, description of the DC conductivities requires taking into account sliding of the background stripes. As it turns out, the slides move at some (small) velocity $v_s$  which is of the same order as the applied electric field. This corresponds roughly to a background solution of the form $f(x-v_st,u)$, so that expanding and small $v_s$ leads to $\delta f(t,x,u) = - v_s t \partial_x f(x,u)$.
These functions no longer satisfy the equations of motion exactly, but motivate us to write down a more general Ansatz: 
\bea \label{Ansatz2}
 \delta a_x(t,x,u) &=& -(E_x - p'(x))t + \delta a_x(x,u) \\ \delta a_t(t,x,u) & =& p(x) + \delta a_t(x,u) - v_s t\, \partial_x a_t(x,u)\ ,\label{Ansatz3}
\eea
and $\delta f(t,x,u) = \delta f(x,u) - v_s t\, \partial_x f(x,u) $ for the other fluctuations.\footnote{We could also include a term of the form $\kappa \partial_x f(x,u)$ in the Ansatz, but the constant $\kappa$ can be canceled by shifts of the coordinate system, so it is irrelevant for the physics.} Since the time dependence multiplies a trivial solution of the fluctuation equations, the terms linear in time cancel in the fluctuation equations, and we are left with a set of consistent, time independent equations. Numerical comparison to the optical conductivity at small frequencies in 
Sec.~\ref{sec:AC conductivity} will verify that the above Ansatz is correct.
The auxiliary function $p(x)$ will also cancel 
thanks to gauge covariance although it will affect the boundary conditions. To pin down the separation of $\delta a_t(t,x,u)$ into $p(x)$ and $\delta a_t(x,u)$, we require that $p(x)$ be periodic with mean zero. For more discussion on this and the associated gauge choice, see App.~\ref{app:gauge}.

For the boundary conditions, we require that all (gauge invariant) source terms vanish  in the UV (expect for $E_x$ which was written explicitly above). That is, $\partial_u\delta \psi(x,0)=0=\delta \psi(x,0)$ and $\delta f(x,0)=0$ for the other functions.  In the IR, we assume the usual ingoing boundary conditions at the horizon. In the case of DC conductivity, this means that the fluctuations are regular after switching from $t$ to the ingoing Eddington-Finkelstein (tortoise) coordinate $v= t + 1/4 \log(1-u) + \cdots$ so that all components of the metric are nonsingular.
Because the background is regular at the horizon and the leading source terms are removed by the $x$-derivative, the boundary conditions for the fluctuations are independent of $v_s$. The fluctuation equations themselves are, however, modified: $v_s$ acts effectively as an extra source in these equations. Regularity or boundary conditions will not set the value of $v_s$, so it remains unfixed in the DC analysis. It turns out that we can fix the value of the speed $v_s$ by considering the DC limit of the optical conductivities as discussed in Sec.~\ref{sec:AC conductivity}.  

The calculation of the DC conductivity in terms of horizon quantities follows the standard recipe, see, {\emph{e.g.}}, \cite{Donos:2014uba}. As pointed out above, the idea is to identify a conserved bulk current which can be used to write several expressions in terms of horizon data. We begin by studying the fluctuation equation for $\delta a_x$ and the constraint (equation for $\delta a_u$), which give us the equations
\be \label{curreqsvs}
 \partial_u (G_1\ \partial_u \delta a_x(x,u)+v_s G_2) = v_s K_1\ , \qquad \partial_x (G_1\ \partial_u \delta a_x(x,u)+v_s G_2) = v_s K_2 \ ,
\ee
where the functions $G_i$, and $K_i$ are complicated functions which depend only on the background.  
Consistency of these equations requires $\partial_x K_1 = \partial_u K_2$.  
Actually one can verify that
$K_1 = \partial_u K$ and $K_2 =\partial_x K$ for a known function $K$ (see Appendix~\ref{app:current}). 
Consequently,~\eqref{curreqsvs} can be written in a form 
\be \label{modconscurr}
 \partial_u (G_1\ \partial_u \delta a_x(x,u)+v_s \tilde G_2) = 0 = \partial_x (G_1\ \partial_u \delta a_x(x,u)+v_s \tilde G_2)
\ee
where $\tilde G_2 = G_2 - K$. 
From here we immediately deduce that the combination $G_1\ \partial_u \delta a_x(x,u)+v_s \tilde G_2$ is constant, {\emph{i.e.}}, the conserved bulk quantity that we sought for.
The boundary value at the UV is given by
\be \label{conservcondbdry}
\frac{1}{\sqrt{2}} \lim_{u\to 0} \left(G_1\ \partial_u \delta a_x(x,u)+v_s \tilde G_2\right) = j_x(x) - v_s d(x) \ ,
\ee
where $d(x) = -\partial_u a_t(x,0)$. Therefore, the modulation of the current is given by the movement of the charge density in the stripes. 

We now proceed by computing the conserved current in the IR. For this we need to expand the background solution in a power series close to the horizon.
Near the horizon we expect that 
\be \label{deltaaxhor}
 \delta a_x(x,u) = \delta a_{x,0}(x) \log(1-u)+\morder{(1-u)^0}\ .
\ee
Further expanding the background solution as
\begin{align} \label{bghorexp1}
 \psi(x,u) &= \psi_0(x) + \morder{1-u}\ ,& \qquad  z(x,u) &= z_0(x) + \morder{1-u}\ , & \\
  a_y(x,u) &= a_{y,0}(x) + \morder{1-u}\ ,&\qquad a_t(x,u) &= a_{t,0}(x)(u-1) +\morder{(1-u)^2}
 \label{bghorexp2}
\end{align}
and inserting in the expression for $G_1$ and $\tilde G_2$ we find that
\be\label{conservcond}
\frac{1}{\sqrt{2}}  \lim_{u\to 1} \left(G_1\ \partial_u \delta a_x(x,u)+v_s \tilde G_2\right) = - \left( 4 \delta a_{x,0}(x) -v_s a_{t,0}(x) \right)\hat \sigma(x) \ ,
\ee
where $\hat \sigma(x)$ is given in~\eqref{localcond1}. The horizon expansion in terms of the ingoing Eddington-Finkelstein coordinate
\be
 \delta a_x(t,x,u) = -(E_x - p'(x))\,v + \left[\frac 14 E_x - \frac 14 p'(x)+ \delta a_{x,0}(x)\right] \log(1-u)+\morder{(1-u)^0}
\ee
is regular if
\be \label{regcond}
 \delta a_{x,0}(x) = -\frac 14 \left(E_x - p'(x)\right)\ .
\ee
Combining~\eqref{conservcondbdry},~\eqref{conservcond}, and~\eqref{regcond} gives finally the expression for the longitudinal current
\be\label{jceq}
 j_x(x) - v_s d(x) = j_c = \left(E_x -p'(x)+ v_s a_{t,0}(x) \right)\hat \sigma(x) \ ,
\ee
where $j_c$ is a constant.
The remaining task is to eliminate the auxiliary function $p(x)$. 
This can be done by taking a spatial average (\ref{eq:average}) of the latter part of (\ref{jceq}), after first dividing by $\hat\sigma$,  to obtain
\be\label{jccond}
 j_c\langle \hat\sigma^{-1}\rangle = E_x+v_s\langle a_{t,0}\rangle \ .
\ee
We are now equipped to spell out the final result for the longitudinal DC conductivity; after collecting all the results,
\be \label{sxxfinal1}
 \sigma_{xx}(x) = \frac{j_x(x)}{E_x} = \frac{v_s}{E_x} d(x) + \frac{j_c}{E_x} =\frac{v_s}{E_x} d(x)+ \langle\hat \sigma^{-1} \rangle^{-1}\left(1+\frac{v_s}{E_x}\langle a_{t,0}\rangle \right) \ .
\ee
The modulation of the local chemical potential obeys
\be
 p'(x) = E_x + v_s a_{t,0}(x) -\frac{j_c}{\hat \sigma(x)} = E_x\left(1-\frac{\langle\hat \sigma^{-1} \rangle^{-1}}{\hat \sigma(x)}\right) + v_s\left( a_{t,0}(x) - \langle a_{t,0}\rangle \frac{\langle\hat \sigma^{-1} \rangle^{-1}}{\hat \sigma(x)}\right) \ .
\ee

We can gain additional insight by rearranging the terms in~\eqref{sxxfinal1}. Namely, by using the identity~\eqref{chargeidentity}, the conductivity becomes
\begin{align}\label{eq:sigmaxx}
  \sigma_{xx}(x) &= \langle\hat \sigma^{-1} \rangle^{-1} +  \frac{v_s}{E_x}\left(d(x)-\langle d\rangle\right) +\frac{\sqrt{2}v_s}{E_x} \langle c(\psi_0) a_{y,0}'\rangle & \\\nonumber
 & -\frac{v_s}{E_x}\left\langle a_{t,0}\hat \sigma \left(a_{y,0}'(x)^2+\psi_0'(x)^2+z_0'(x)^2\right)\right\rangle +\frac{v_s}{E_x}\left(\langle a_{t,0}\rangle\langle\hat \sigma^{-1} \rangle^{-1} -\langle a_{t,0}\hat \sigma\rangle \right) \ . &
\end{align}
Let us interpret the result just obtained. We notice that the right-hand-side of (\ref{eq:sigmaxx}) is a sum of distinct terms and suggests that there are several different mechanisms that contribute to the total conductivity. First of all, the first term on the RHS is the familiar contribution that one would get from classical circuit of resistors in a series: Think momentarily the stripes as a static configuration with given $x$ contributing differently and additively to the total resistance. Second, the only $x$-dependent term in (\ref{eq:sigmaxx}) is the second one on the first row: $\propto d(x)-\langle d\rangle$. This is nothing but the modulation of the total charge density of the system and can be viewed as the CDW. The next two terms have, as their most dominant pieces, either the spatial derivative of the scalar $\psi$ or the transverse gauge field $a_y$, evaluated at the horizon. This is directly related to the SDW of the system. It is thus no surprise that as the background in mostly in the SDW phase and not so much in CDW, this fact is also signaled in the conductivities. Finally, the last term on the RHS of (\ref{eq:sigmaxx}) is a small residual of the spatial averaging over the stripes. One can check that the numerical value of this difference is minuscule and vanishes in the limit of $\mu\to \mu_c$ where the stripe solution continuously approaches the homogeneous solution; for the homogeneous solution this term is identically zero.

As a final step, we can perform the spatial averaging of (\ref{eq:sigmaxx}). This simply removes the term proportional to the modulation of the charge density:
\begin{align}\label{eq:sigmaxxave}
  \langle\sigma_{xx}\rangle &= \langle\hat \sigma^{-1} \rangle^{-1} +\frac{\sqrt{2}v_s}{E_x} \langle c(\psi_0) a_{y,0}'\rangle &\\\nonumber 
  &-\frac{v_s}{E_x}\left\langle a_{t,0}\hat \sigma \left(a_{y,0}'(x)^2+\psi_0'(x)^2+z_0'(x)^2\right)\right\rangle +\frac{v_s}{E_x}\left(\langle a_{t,0}\rangle\langle\hat \sigma^{-1} \rangle^{-1} -\langle a_{t,0}\hat \sigma\rangle \right) \ .  &
\end{align}

\subsection{Adding electric field and current parallel to stripes}\label{sec:jyandEy}

Thus far, we have only discussed the longitudinal conductivity $\sigma_{xx}$ from turning on an electric field perpendicular to the stripes. We also wish to understand the response in the other direction and investigate the associated Hall conductivity $\sigma_{yx}$. In order to compute the transverse current and remaining components of the DC conductivity we need to generalize the previous analysis. 
It turns out that with little effort we can also allow a constant electric field (and current) in the $y$-direction and a slightly modified analysis can be carried out. 

When the electric field is parallel to the stripes, we expect that the stripes stay still, since there is no preferred direction for their motion. But as we want to analyze $\sigma_{yx}$ also, we consider the possibility of nonzero $v_s$.
That is, we write the Ansatz\footnote{Actually it turns out that terms depending on $v_s$ will cancel due to parity when the electric field is in the $y$-direction. Therefore the DC conductivities are independent of $v_s$, but one can check that indeed $v_s=0$ by using the numerical solutions of Sec.~\ref{sec:AC conductivity}.}
\be \label{Ansatzay}
 \delta a_y(t,x,u) = - E_y t +\delta a_y(x,u) - v_s t \partial_x a_y(x,u)
\ee
and keep other fields as above in (\ref{Ansatz2})-(\ref{Ansatz3}), so that $E_x$ is included as well. 
Then the fluctuation of the current is given by
\be \label{jyDC}
 j_y(x) =  \partial_u \delta a_y(t,x,u=0) = \partial_u \delta a_y(x,u=0) - v_s t J_y'(x) \equiv \bar j_y(x) - v_s t J_y'(x)\,.
\ee
The second  term, which is linear in $t$, is the contribution to the current fluctuation due to the background current $J_y(x)$ moving together with the stripes. Notice that the total current is $J_y(x)+j_y(x)$, where the terms involving the background current can be rearranged to $J_y(x)-  v_s t J_y'(x) \simeq J_y(x- v_s t)$.

The fluctuation of the current is ambiguous: we could have, for example, chosen the origin of time differently, effectively replacing $t \mapsto (t-t_0)$ in~\eqref{jyDC}. This ambiguity corresponds to an \order{v_s} shift in the $x$-direction of the background solution with respect to which the fluctuation is defined or equivalently adding a term proportional to the zero mode (the $x$-derivative of the background) to the fluctuations. For the averaged conductivities we discuss here, the divergence and the ambiguity are, however, absent because the terms involving the background current average to zero; $\langle J_y' \rangle =0$. The divergence and the ambiguity can be resolved in the AC case, as we will discuss in Sec.~\ref{sec:AC conductivity}.

The conserved current in~\eqref{conservcond} is modified to
\be
G_1\ \partial_u \delta a_x(x,u)+v_s \tilde G_2 \mapsto G_1\ \partial_u \delta a_x(x,u)+v_s \tilde G_2 + E_y G_3 \ ,
\ee
where $G_3$ depends only on the background. In particular, it contains a term $\sim 2c(\psi(x,u))$  arising from the Chern-Simons action. 
Using this current, we find that 
\bea
 &&\frac{1}{\sqrt{2}}\lim_{u\to 1} \left(G_1\ \partial_u \delta a_x(x,u)+v_s \tilde G_2 + E_y G_3\right) = \sqrt{2} E_y c(\psi_0(x)) \nn\\
 &&\qquad \qquad-  \left( 4 \delta a_{x,0}(x) + E_y a_{t,0}(x) a_{y,0}'(x) -v_s a_{t,0}(x) \right)\hat \sigma(x) \ ,
\eea
where the horizon values of the background and the fluctuations were defined in~\eqref{localcond1} and \eqref{deltaaxhor}--\eqref{bghorexp2}. Eq.~\eqref{jceq} generalizes to
\be
 j_x(x) - v_s d(x) = j_c = \sqrt{2} E_y c(\psi_0(x)) + \left(E_x -p'(x)+ v_s a_{t,0}(x)- E_y a_{t,0}(x) a_{y,0}'(x) \right)\hat \sigma(x) 
\ee
The parity symmetry~\eqref{symmbgparity} of the background solutions, which have zero background magnetic field,
implies that
\be
 \langle c(\psi_0) \hat \sigma^{-1} \rangle = 0 = \langle a_{t,0} a_{y,0}' \rangle \,.
\ee
Consequently the condition~\eqref{jccond} is unchanged, and the current is therefore independent of $E_y$. That is,
\be
\sigma_{xy}(x) = 0\,.
\ee
The generic expression for $p'(x)$ becomes
\bea \label{pprimegen}
 p'(x) &=& E_x\left(1-\frac{\langle\hat \sigma^{-1} \rangle^{-1}}{\hat \sigma(x)}\right) + v_s\left( a_{t,0}(x) - \langle a_{t,0}\rangle \frac{\langle\hat \sigma^{-1} \rangle^{-1}}{\hat \sigma(x)}\right)\nn\\
 &&+\left(\frac{\sqrt{2}c(\psi_0(x))}{\hat \sigma(x)} - a_{t,0}(x) a_{y,0}'(x) \right) E_y \,.
\eea

There is no conserved current corresponding to $a_y$, but because the action depends on $\delta a_y$ only through its derivatives,
\be
\left\langle\frac{\delta S}{\delta ( \partial_u \delta a_y(t,x,y) )} \right\rangle
\ee
is independent of $u$. The UV limit is given by
\be \label{jyuto0}
 \lim_{u \to 0}\frac{\delta S}{\delta ( \partial_u \delta a_y(t,x,y) )}  = \sqrt{2} \bar j_y(x) \,.
\ee
In order to compute the IR limit, we need the IR expansions for all fluctuations (expect for $\delta a_x$). We write\footnote{Notice that the term $-p(x)$ here cancels a similar term in~\eqref{Ansatz3} so that the full fluctuation $\delta a_{t}(t,x,u)$ vanishes at the horizon, which is expected because the background field $a_t$ also vanishes.}
\be
 \delta a_{t}(x,u) = -p(x) + \morder{1-u}
\ee
and
\be
 \delta f(x,u) = \delta f_0(x) \log(1-u) + \morder{(1-u)^0}
\ee
for the other fields. Then the IR limit becomes
\bea \label{jyuto1}
 &&\lim_{u \to 1} 
 \frac{\delta S}{\delta ( \partial_u \delta a_y(t,x,y) )} 
 = -2 c(\psi_0(x))\left(E_x - p'(x)\right) \\\nn
&& +\sqrt{2} \hat \sigma(x) \big[ - 4 (1+z_0'(x)^2+\psi_0'(x)^2) \delta a_{y,0}(x) + (E_x-p'(x))a_{t,0}(x) a_{y,0}'(x) \\\nn
&& + 4 a_{y,0}(x) z_0(x) \delta z_0(x)+ 4 a_{y,0}(x) \psi_0(x) \delta \psi_0(x)\big] \ .
\eea
The regularity conditions at the horizon, which generalize~\eqref{regcond}, arise from the terms proportional to $t$ in our Ansatz,
\be
 \delta a_{y,0}(x) = -\frac{E_y}{4} - \frac{v_s}{4} a_{y,0}'(x) \,,\quad \delta \psi_0(x) = -\frac{v_s}{4} \psi_0'(x)\,,\quad \delta z_0(x) = -\frac{v_s}{4} z_0'(x) \ .
\ee
Inserting these and $p'(x)$ from~\eqref{pprimegen} in~\eqref{jyuto1}, taking the average over $x$, and equating with the UV limit~\eqref{jyuto0} we find
\be
 \langle j_y\rangle = \langle \bar j_y\rangle = \left\langle \hat \sigma(1+z_0'^2+\psi_0'^2) + \frac{1}{\hat \sigma}\left(\sqrt{2}c(\psi_0) - \hat \sigma a_{t,0} a_{y,0}'\right)^2 \right\rangle E_y \ .
\ee
Thanks to parity symmetry, the terms proportional to $E_x$ and $v_s$ vanish when taking the average. The final results for the averaged conductivities therefore become
\bea
 \langle\sigma_{yy}\rangle &=& \left\langle \hat \sigma(1+z_0'^2+\psi_0'^2) + \frac{1}{\hat \sigma}\left(\sqrt{2}c(\psi_0) - \hat \sigma a_{t,0} a_{y,0}'\right)^2 \right\rangle \label{eq:sigmayyave}\\
 \langle\sigma_{yx}\rangle &=& 0 \,.\label{eq:sigmayxave}
\eea

The first term in the expression for $\langle\sigma_{yy}\rangle$, i.e, $\hat\sigma$, can be interpreted as the result for classical circuits in the same way as the first term for $\sigma_{xx}$ in~\eqref{eq:sigmaxx}, except that the classical resistors are now in parallel rather than in series. This expression is, however, enhanced by additional derivative terms $\propto z_0'^2+\psi_0'^2$ due to the stripes. It turns out the last term in~\eqref{eq:sigmayyave} is typically the dominant term. As it includes $c(\psi_0)$ and $a_{y,0}'$, it is mostly induced by the SDW component of the background.
We will see in the numerical analysis of Sec.~\ref{sec:AC conductivity} that the values of $ \langle\sigma_{yy}\rangle$ and $ \langle\sigma_{xx}\rangle$ are almost equal independently of the value of $\mu$ (see Fig.~\ref{fig:DCvsmudep}). We speculate that this is due to the dominance of the SDW (see App.~\ref{app:fluct}.)

\subsection{Summary}

We found:
\begin{itemize}
 \item The longitudinal conductivity $\sigma_{xx}(x)$ could be written in terms of horizon data, and it naturally splits in to contributions from different sources (\ref{eq:sigmaxx}).
 \item There is no conserved (local, $x$-dependent) quantity associated with the current in the $y$-direction.
 \begin{itemize}
  \item Only the averaged conductivity $\langle\sigma_{yy}\rangle$ could be written in terms of horizon data (\ref{eq:sigmayyave}).
  \item The averaged Hall conductivity $\langle\sigma_{yx}\rangle$ was found to vanish (\ref{eq:sigmayxave}). Locally $\sigma_{yx}(x)$ diverges, a phenomenon which will be addressed in Sec.~\ref{sec:AC conductivity}.
 \end{itemize}
 \item As a direct result of parity symmetry,  $\sigma_{xy}(x)=0$.
\end{itemize}

We now turn to studying optical, or alternating, currents numerically and show that their zero-frequency limit  precisely match the (semi-)analytic results that we obtained in this section.

\section{Optical conductivities}
\label{sec:AC conductivity}

\subsection{Setup and symmetries}

In order to compute the optical conductivities, we carry out the analysis of fluctuations for the action~\eqref{totalact} on top of the modulated backgrounds constructed in~\cite{Jokela:2014dba}. To do so, we separate the fields $\psi$, $z$, $a_t$, $a_y$, and $a_x$ into the background and fluctuation components: 
\be 
 f = \bar f(x,u) + \delta \hat f(t,x,u)
\ee 
for each field $f$ (with the understanding that $\bar a_x(x,u)$ vanishes everywhere). For the fluctuations, we further write the Ansatz
\be
 \delta \hat f(t,x,u)= e^{-i\omega t} \delta \hat f(x,u) = e^{-i\omega t} (1-u)^{-i\omega/4} \delta f(x,u) \ ,
\ee
where the nonanalytic term at the horizon $u=1$ has been separated  in the last expression, so that the infalling modes $\delta f(x,u)$ are regular on the horizon. We have chosen the gauge $\delta a_u = 0$. The analysis is similar to that of~\cite{Araujo:2015hna} where optical conductivities were analyzed in the presence of a lower dimensional interface.

We restrict ourselves to the backgrounds with vanishing magnetic field and quark mass here. In order to turn on (infinitesimal) electric fields in the $x$ and $y$ directions, we choose the following boundary conditions:
\begin{align} \label{Exbc}
 i \omega \delta a_x (x,0) + \partial_x \delta a_t(x,0) &= i \omega\, e_x & \\
 \delta a_y (x,0) &= e_y \ . & \label{Eybc} 
\end{align}
As usual, an extra factor of $\omega$ was included in the amplitudes of the electric fields so that the physical electric fields are $\propto i\omega\, e_{x,y} e^{-i\omega t}$. The fluctuation equations also impose charge conservation on the boundary, which we choose to implement as an explicit boundary condition:
\be
 \partial_t \partial_u \delta \hat a_t(t,x,0) = \partial_x\partial_u \delta \hat a_x(t,x,0) \ .
\ee
No other sources are turned on, so that
\be
 \partial_u \delta \psi(x,0) = 0 = \delta z(x,0) \ .
\ee
In addition, we require infalling conditions, i.e., that $\delta f$ are regular at the horizon, and that $\delta a_t(x,1) = 0$.

As discussed above, the striped background solutions realize two discrete symmetries of the action, rotation by $\pi$~\eqref{symmbg180rot} and parity~\eqref{symmbgparity}. 
These symmetries also show up in the fluctuation analysis. There is, however, an important difference with respect to the symmetry of the background. For example, $\delta a_y$ cannot be odd and  $\delta a_t$ cannot be even as one would expect for rotations by $\pi$ because that would violate the boundary conditions~\eqref{Eybc} and~\eqref{Exbc}, respectively. This suggests that this symmetry is broken. The symmetry is, however, preserved because the Lagrangian for the fluctuations is by definition quadratic, which allows us to insert an extra sign in the transformation of each field. After this, $\delta a_y$ and $\delta a_x$ are even while the other fluctuations are odd under the reflection $x \mapsto -x$, corresponding to the rotation by $\pi$, which is consistent with the boundary conditions.
As for the parity symmetry, the transformations of the fields are either the same or with opposite signs with respect to the transformations of the background, depending on whether we have turned on $e_x$ or $e_y$, because only $e_x$ breaks parity (and with the understanding that the vanishing background field $\bar a_x$ is odd). All these symmetries are summarized in Table~\ref{tab:symmetries}.
 
\begin{table}[!ht]
\begin{center}
\begin{tabular}{ lcl || c | c | c | c | c}
\multicolumn{3}{ l || }{symmetry} & $\psi$ & $z$ & $a_t$ & $a_y$ & $a_x$ \\
\hline \hline
b.g. & $\pi$-rotation & & $+$  & $+$ & $+$ & $-$ & n/a \\
b.g. & $\mathcal{P}$ & & $-$  & $+$ & $+$ & $+$ & n/a  \\
\hline
fluct. & $\pi$-rotation & & $-$  & $-$ & $-$ & $+$ & $+$ \\
fluct. & $\mathcal{P}$ & ($e_x$) & $+$  & $-$ & $-$ & $-$ & $+$ \\
fluct. & $\mathcal{P}$ & ($e_y$) & $-$  & $+$ & $+$ & $+$ & $-$ \\
\end{tabular}
\caption{Summary of the discrete symmetries (rotation by 180 degrees and parity) of the striped background (b.g.) and its fluctuations (fluct.) for $b=0=m$.  
Notice that it is actually the background field $\psi-\pi/4$ which is odd under parity symmetry, not $\psi$ itself. The behavior of the fluctuations under parity depends on the direction of the electric field as noted in the table.} 
\label{tab:symmetries}
\end{center}
\end{table} 
 
We have implemented the first of the discrete symmetries, which is independent of the direction of the electric field, in the numerical solutions of the fluctuations. Just as for the background, implementing this symmetry of the fluctuations allows us to reduce the domain of the solutions from $x= 0 \ldots L$, where $L$ is the periodicity of the whole solution, to $x = 0 \ldots L/2$. For a background  depending on $x$ but not on $y$, the rotation by 180 degrees effectively becomes a reflection, the axis of which was chosen to be at $x=0$ (then the axis of reflection for the parity symmetry is at $x=L/4$). 

We  discretize the fluctuation equations on a suitably chosen grid. The points are taken to be evenly spaced grid in the $x$-direction and from a Gauss-Lobatto grid in the $u$-direction. The equations are then solved by using a pseudospectral approximation for the derivatives which also takes into account the symmetry of the functions (as was done for the background in~\cite{Jokela:2014dba}). We used about 40 grid points in each direction for the numerical results in this article. From the solution, we then extract the conductivities as
\be
 \left(\begin{array}{c}
        j_x(\omega,x) \\ 
        j_y(\omega,x)
       \end{array}
\right)
 =  \left(\begin{array}{c}
        \partial_u \delta \hat a_x(x,u=0)\\ 
        \partial_u \delta \hat a_y(x,u=0)
       \end{array}
\right)
 =\  \left(\begin{array}{cc}
        \sigma_{xx}(\omega,x) & \sigma_{xy}(\omega,x) \\ 
        \sigma_{yx}(\omega,x) & \sigma_{yy}(\omega,x)
       \end{array}
\right)\left(\begin{array}{c}
        i\omega\,e_x\\ 
        i\omega\,e_y
       \end{array}\right) \ .
\ee

The current $j_y$ ($j_x$) is odd under parity (see Table~\ref{tab:symmetries}) if $e_x$ ($e_y$) is turned on. Therefore, the nondiagonal components of the optical conductivity vanish when averaged over $x$. The same will no longer be true after turning on a magnetic field, because it destroys the parity symmetry.

\subsection{Limit of small frequency} 
\label{sec:smallfreq}
 
It is instructive to compare the expressions~\eqref{Ansatz2} and~\eqref{Ansatz3} of the DC computation to the $\omega$-dependent fluctuations in the limit $\omega \to 0$. We first sketch some generic features of this limit. To obtain the DC conductivity, we need to turn on a constant infinitesimal electric field by turning on a fluctuation of the gauge field $\delta A = -E t+\cdots$ on the boundary. For the optical conductivity, we solve exactly the same fluctuation equations with a different boundary condition, $\delta A = \hat E e^{-i\omega t}/(i\omega) $, which gives an oscillating electric field $\hat E e^{-i\omega t}$. The electric fields of the DC and AC solutions trivially match if $E = \hat E$ in the limit $\omega \to 0$, but the limit for $\omega$-dependent gauge field is singular. In order to resolve the singularity, we need to subtract a constant from the gauge field and consider instead
\be \label{Aomegasketch}
 \delta \hat A = \frac{\hat E e^{-i\omega t}}{i\omega} - \frac{\hat E}{i\omega} \ \xrightarrow[\omega\to 0]{} \ - \hat E t \ .
\ee
Notice that adding the extra term is acceptable for any value of $\omega$ because it is independent of $t$ and can be therefore gauged away. Choosing $\hat E = E$, the limit agrees with the boundary condition for the DC fluctuations, so the solutions agree as well.\footnote{The convergence of both the boundary conditions and the solutions is only pointwise, not uniform in $t$, but this is sufficient for the comparison.}

Our case is somewhat more complicated than the sketch above because we have a Goldstone mode due to the movement of the stripes and the modulation of the chemical potential in the $x$-direction. In particular, our DC result contained an extra parameter  (the velocity of the stripes $v_s$) due to the Goldstone mode, which was not fixed by the boundary conditions, whereas no such parameter exists at finite $\omega$. 

Motivated by the above discussion, we require that the terms with linear time dependence in the $\omega$ expansions of the AC fluctuations match with the DC fluctuations in~\eqref{Ansatz2},~\eqref{Ansatz3}, and~\eqref{Ansatzay}. This leads to
\begin{align}
\label{axsmallo}
 \delta \hat a_x(t,x,u) &= \frac{E_x-p'(x)+\morder{\omega}}{i\omega}e^{-i\omega t} \; = \frac{E_x-p'(x)}{i\omega} - \left(E_x-p'(x)\right) t +\morder{\omega^0}\\
  \delta \hat a_y(t,x,u) &= \frac{E_y+v_s \partial_x \bar a_y(x,u)+\morder{\omega}}{i\omega}e^{-i\omega t} \nonumber\\ \label{aydiv}
  &= \frac{E_y+v_s \partial_x \bar a_y(x,u)}{i\omega} - \left(E_y+v_s \partial_x \bar a_y(x,u)\right) t +\morder{\omega^0}\\
 \delta \hat f(t,x,u) &= \frac{v_s \partial_x \bar f(x,u) +\morder{\omega}}{i\omega} e^{-i\omega t}= \frac{v_s \partial_x \bar f(x,u)}{i\omega} - v_s t\,  \partial_x \bar f(x,u)+\morder{\omega^0} \ ,
 \label{smallomega}
\end{align}
for the AC fluctuations, where $f = a_t$, $\psi$, $z$ and we only included the time dependent pieces of the \order{\omega^0} terms. 

We now argue that~\eqref{axsmallo}--\eqref{smallomega} are correct. Notice that the divergent terms $\propto 1/\omega$ in the above expansions satisfy the fluctuation equations, and can therefore be subtracted as in~\eqref{Aomegasketch}. These terms include the Goldstone mode (given as the $x$-derivative of the background) and terms of the gauge fields that are independent of $t$ and $u$, which could be gauged away. After the subtraction, the AC fluctuations have a regular $\omega \to 0$ limit. The only nontrivial boundary conditions both for the DC and AC fluctuations are those of the gauge fields $a_x$ and $a_y$. Therefore, comparing~\eqref{axsmallo} and~\eqref{aydiv} to the AC conditions~\eqref{Exbc} and~\eqref{Eybc}, we conclude\footnote{Notice that $\bar a_y(x,0)=0$ and that the $x$-derivatives vanish when averaged over $x$.} that if $E_x = i\omega\, e_x$ and $E_y = i\omega\, e_y$, the boundary conditions and therefore the solutions (with the divergent terms subtracted on the AC side) will match for some value of the speed of the stripes $v_s$. We will use the divergent term in~\eqref{smallomega} to extract this value numerically from the AC fluctuations at small $\omega$.

Noticeably, as seen from~\eqref{aydiv}, if $v_s \ne 0$ the current parallel to the stripes will contain a divergent piece as $\omega \to 0$, which arises from the derivative of the background. That is,
\be \label{jyACdiv}
 j_y(t,x) \equiv \partial_u \delta \hat a_y(t,x,0) = \frac{v_s \partial_x\partial_u \bar a_y(x,0)}{i\omega} +\morder{\omega^0} = \frac{v_s J_y'(x)}{i\omega} +\morder{\omega^0} \ ,
\ee
where $J(y)$ is the modulated current in the background. This reflects similar divergence of the current as $t \to \infty$ in the DC analysis, seen in~\eqref{jyDC}. Since $v_s \propto e_x$, this will result in a modulated divergence of $\mathrm{Im}\,\sigma_{yx}(x)$ as $\omega \to 0$ which will disappear after taking the average over $x$, and which we will study numerically below.

\begin{figure}[!ht]
\center
 \includegraphics[width=0.70\textwidth]{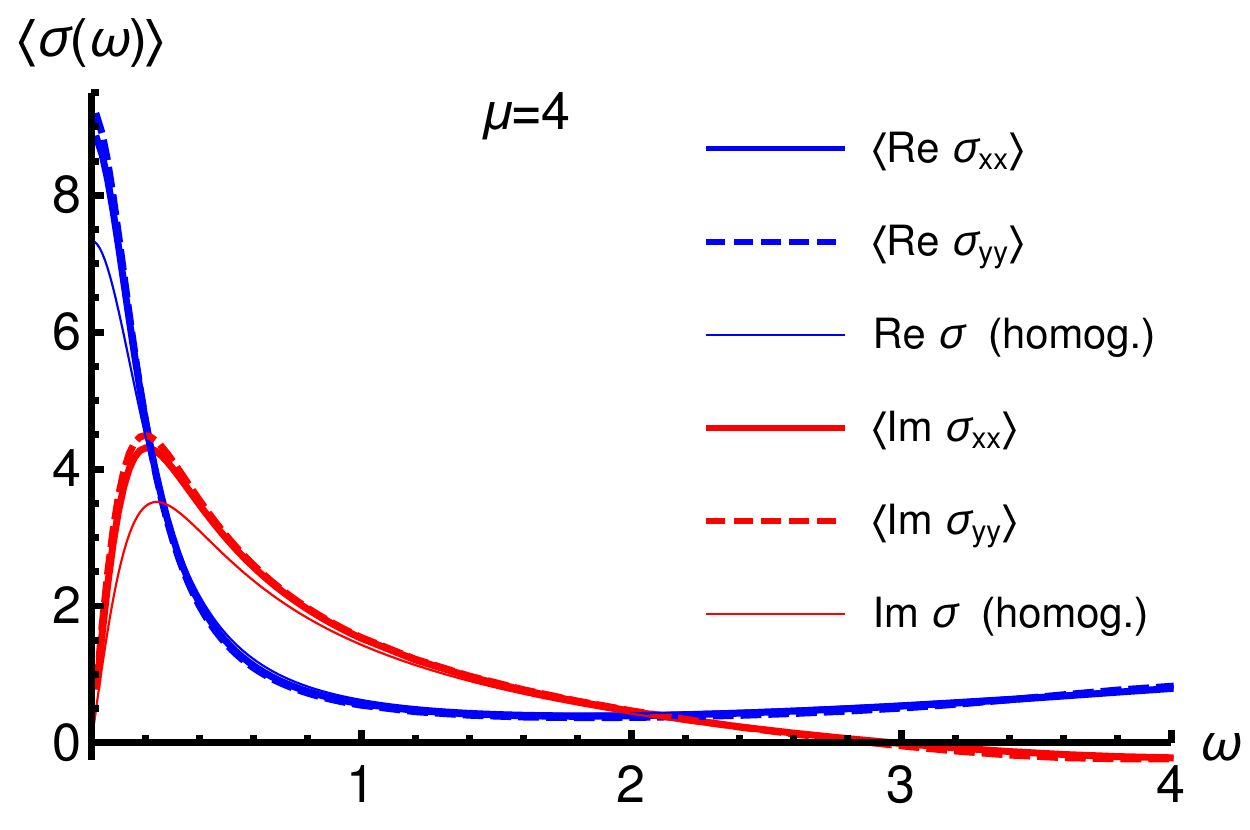}%
  \caption{Optical conductivities for $\mu=4$, averaged over $x$. The solid (dashed) thick curves show the optical conductivity $\sigma_{xx}$ ($\sigma_{yy}$) averaged over $x$ as a function of $\omega$. Real parts are shown in blue and imaginary parts in red. The thin lines are the optical conductivities for the (subdominant) homogeneous background.} 
  \label{fig:conductivitiesmu4}
\end{figure}

\subsection{Numerical results: fixed $\mu$} 

In order to understand the results, it is useful to recall the phase diagram of the model as a function of the (rescaled) chemical potential $\mu$ at zero magnetic field and quark mass. There are two phases, a homogeneous phase at small $\mu$, and a striped phase at large $\mu$, separated by a second-order phase transition at $\mu=\mu_c \simeq 2.66$.

We start by picking a value $\mu=4$ deep in the striped phase. 
The optical conductivities, averaged over the period of $x$, are shown in Fig.~\ref{fig:conductivitiesmu4}. All conductivities show a clear Drude peak at $\omega = 0$ and approach one as $\omega \to \infty$, as expected for conductivities with the standard normalization. The conductivities perpendicular to the stripes ($\sigma_{xx}$, solid lines) and parallel to the stripes ($\sigma_{yy}$, dashed lines) are numerically very close. We argue in App.~\ref{app:fluct} that this may be due to the dominance of the SDW in the striped background. The conductivities in the striped phase are enhanced with respect to those of the (subdominant) homogeneous phase, shown as thin lines.

\begin{figure}[!ht]
\center
 \includegraphics[width=0.70\textwidth]{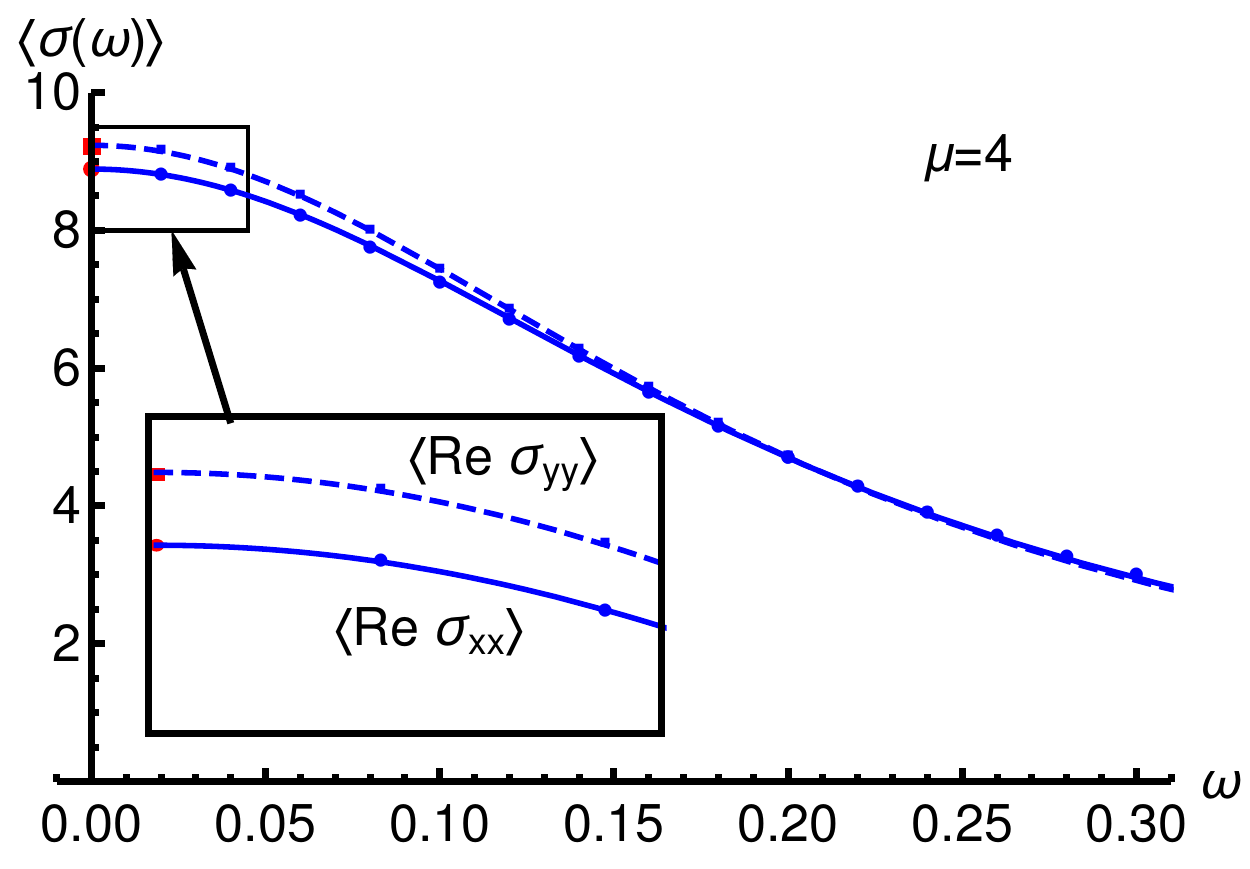}
  \caption{Fit of the optical conductivities using the Drude form, and comparison of the $\omega \to 0$ limit of the optical conductivities to the DC conductivities. The blue dots at finite $\omega$ are our data for the real parts of the optical conductivities and the solid and dashed curves are given by two parameter fits using the Drude form for $\sigma_{xx}$ and $\sigma_{yy}$, respectively. The red dots at $\omega = 0$ are the DC conductivities computed using the formulae of Sec.~\ref{sec:DC conductivity}.} 
  \label{fig:conductivitiesDClimit}
\end{figure}

In Fig.~\ref{fig:conductivitiesDClimit} we fit the data for (the real parts of) the optical conductivities using the Drude formula, i.e.,
\be \label{Drudepeak}
 \sigma(\omega) = \frac{\sigma_0}{1-i\omega \tau}
\ee
and compare the result to DC conductivities from Sec.~\ref{sec:DC conductivity}. The blue dots are our data, and the curves are fits of~\eqref{Drudepeak} to ten data points for $\langle\mathrm{Re}\,\sigma\rangle$ with lowest $\omega$. The speed of the stripe $v_s$, needed to compute the DC conductivity perpendicular to the stripes, was obtained by comparing the fluctuation $\delta \psi$ to $\partial_x \bar\psi$ and using the formula~\eqref{smallomega}. The results for the DC conductivities are shown as the red dots at $\omega=0$.
Explicitly, the fit results for the optical conductivity and the DC conductivity are, perpendicular to the stripes
\be
 \sigma_{0,x} \simeq 8.8901 \ , \quad \sigma_{xx}^\mathrm{DC} \simeq 8.8900 \ ,
\ee
and parallel to the stripes
\be
 \sigma_{0,y} \simeq 9.2348 \ , \qquad \sigma_{yy}^\mathrm{DC} \simeq 9.2266 \ ,
\ee
showing good agreement. 

We have also checked that the sum rule~\cite{Mas:2010ug}
\be
 \lim_{\Lambda \to \infty}\ 2 \int_0^\Lambda d\omega\ \left[ \mathrm{Re}\, \langle \sigma (\omega)\rangle - \mathrm{Re}\, \langle \sigma (\Lambda)\rangle \right] = 0
\ee
is satisfied, within the precision of our numerical solutions, for the averaged conductivity of the (subdominant) homogeneous background as well as to rough degree for the conductivities $\langle \sigma_{xx}\rangle$ and $\langle \sigma_{yy}\rangle$ of the striped background.

\begin{figure}[!ht]
\center
 \includegraphics[width=0.50\textwidth]{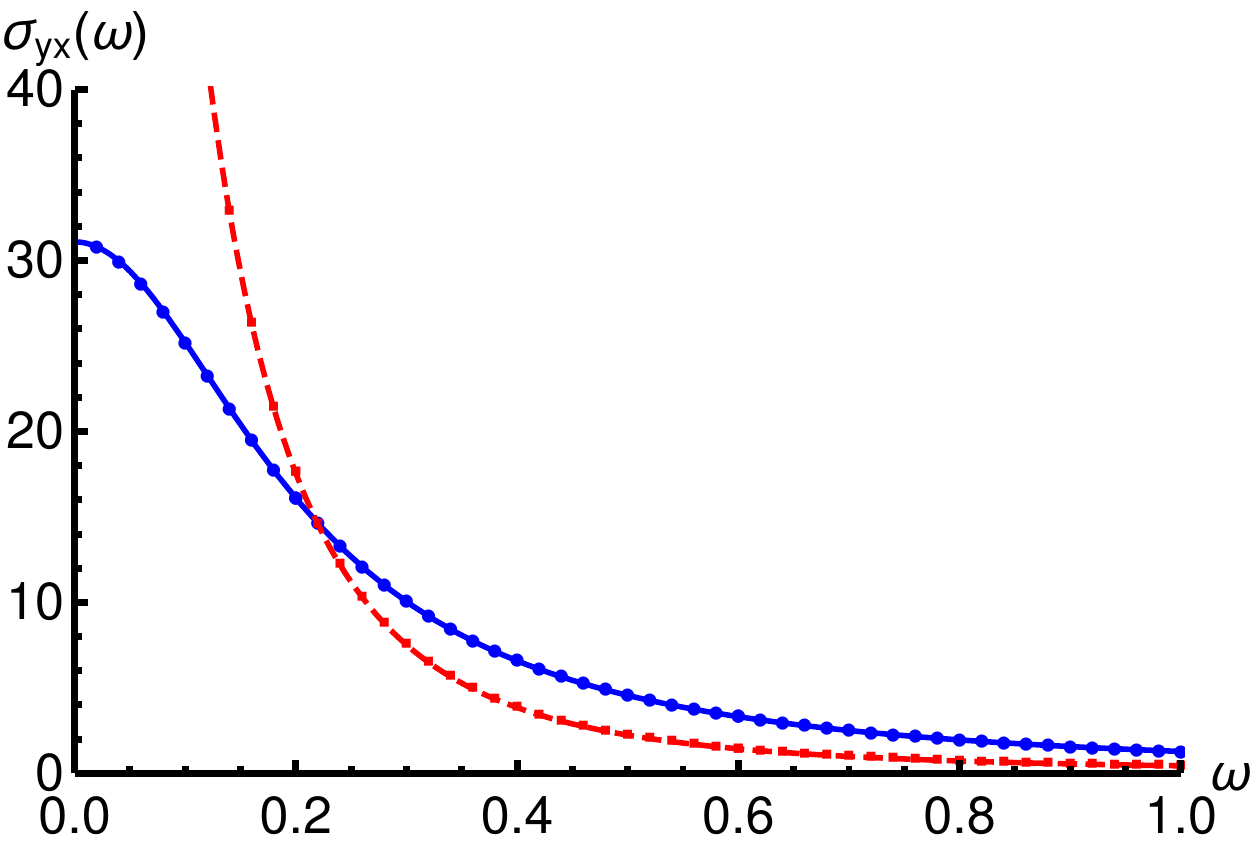}%
 \includegraphics[width=0.50\textwidth]{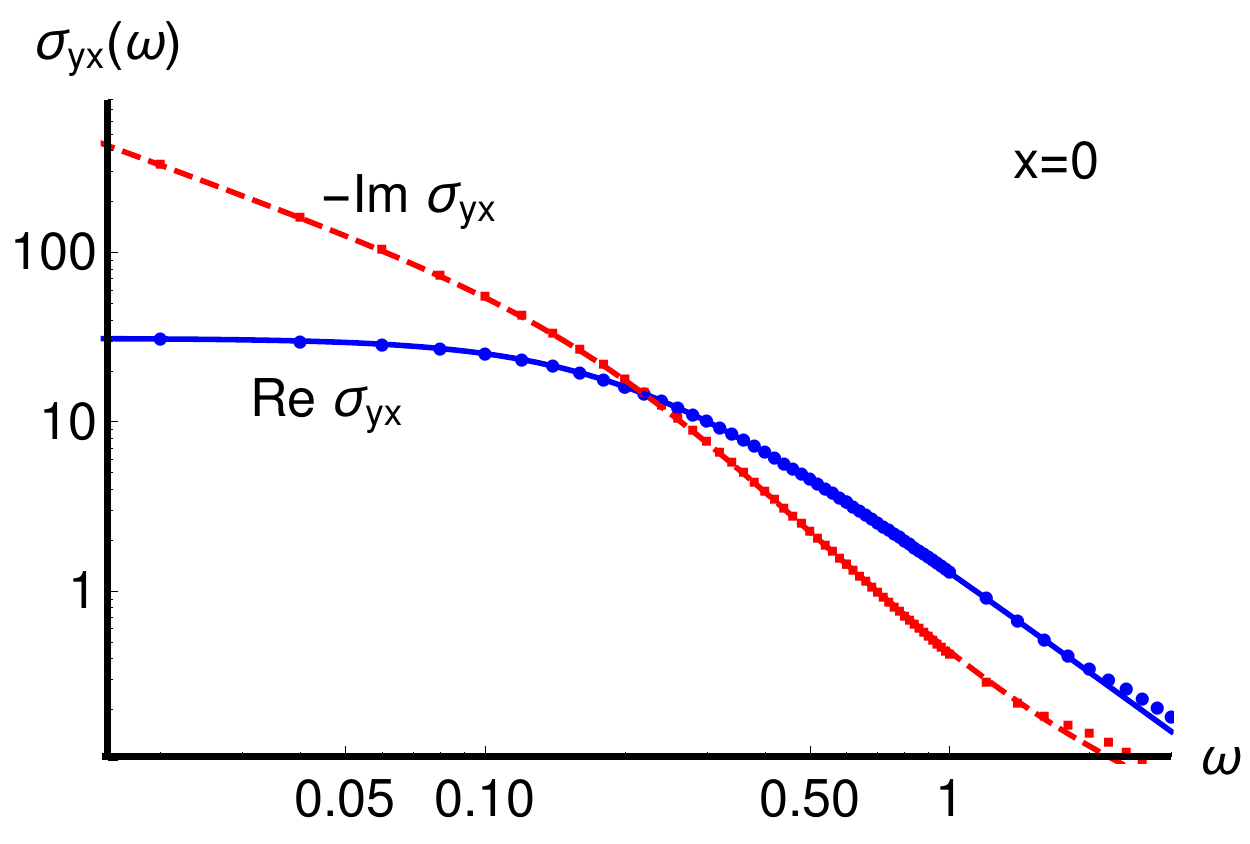}
  \caption{Dependence of the (non-averaged) $\sigma_{yx}$ on $\omega$ at $x=0$. Dots are numerical data and the curves are given by a fit to the formula~\eqref{sigmayxformula}.  Left: linear scale. Right: log-log scale.} 
  \label{fig:sigmayx}
\end{figure}

We have also studied numerically the divergence of the nondiagonal optical conductivity $\sigma_{yx}$ as $\omega \to 0$. As shown above in \eqref{jyACdiv}, the divergence appears in the imaginary part of the conductivity. As it turns out, the $\omega $ dependence of $\sigma_{yx}$ is well described by a Drude peak plus a delta function pointwise in $x$ (recall that the $x$-average vanishes, $\langle \sigma_{yx} \rangle = 0$, for all values of $\omega$). Explicitly, we find that 
\be \label{sigmayxformula}
 \sigma_{yx}(x,\omega) \simeq \frac{\sigma_0(x)}{1-i \omega\, \tau(x)} - \frac{i K(x)}{\omega} = \frac{\sigma_0(x)}{1+\omega^2 \tau(x)^2} + i\left[\frac{\omega\, \tau(x) \sigma_0(x)}{1+\omega^2 \tau(x)^2}- \frac{K(x)}{\omega}\right] 
\ee
at small, finite $\omega$. Comparing to~\eqref{jyACdiv} we see that $K(x) = v_s J_y'(x)$. It would be interesting to understand the appearance of a delta function in the hydrodynamic language of~\cite{Delacretaz:2016ivq}.

In Fig.~\ref{fig:sigmayx} we show $\sigma_{yx}$ as a function of $\omega$ at fixed $x=0$, where the conductivity takes its extremal value, compared to a fit using the formula~\eqref{sigmayxformula}. 
At $x=0$, a least squares fit to 10 data points with lowest $\omega$ gives
\be
 \sigma_0 \simeq 31.089 \ , \qquad \tau \simeq 4.8240 \ , \qquad K \simeq 6.6182 \ .
\ee
The fit is very good up to $\omega \simeq 1$.\footnote{Computing $K$ from the formula $K= v_s J_y'(0)$, with $v_s$ extracted from the divergence of the modulation of $\psi$ as $\omega \to 0$, we find that $K\simeq 6.625$.} Interestingly, the rough cancellation of the two terms in the imaginary part of~\eqref{sigmayxformula}, mimics an intermediate scaling law $\mathrm{Im}\,\sigma_{yx} \sim \omega^{-2.5}$ for $0.2 \lesssim \omega \lesssim 1$, as seen from the red dashed curve in Fig.~\ref{fig:sigmayx} (right).

As a function of $x$ we find that $\tau$ is constant, up to possibly a modulation suppressed by the factor $\sim 10^{-4}$, which cannot be reliably extracted due to limited numerical precision. The same holds for the ratio $K/\sigma_0$. This leaves us with the $x$-dependence of the overall constant, which is roughly proportional to $\cos(2 \pi x/L)$.

\begin{figure}[!ht]
\center
 \includegraphics[width=0.50\textwidth]{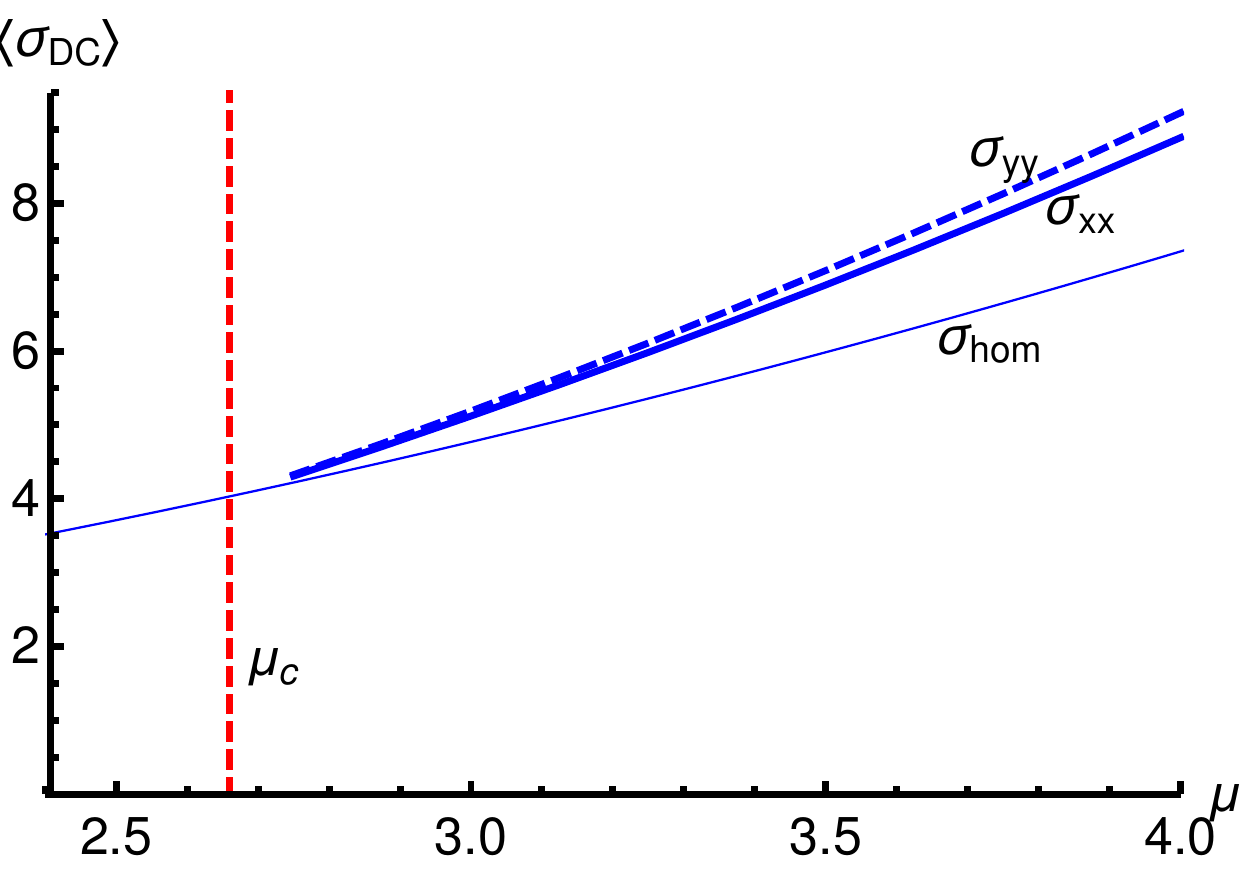}%
 \includegraphics[width=0.50\textwidth]{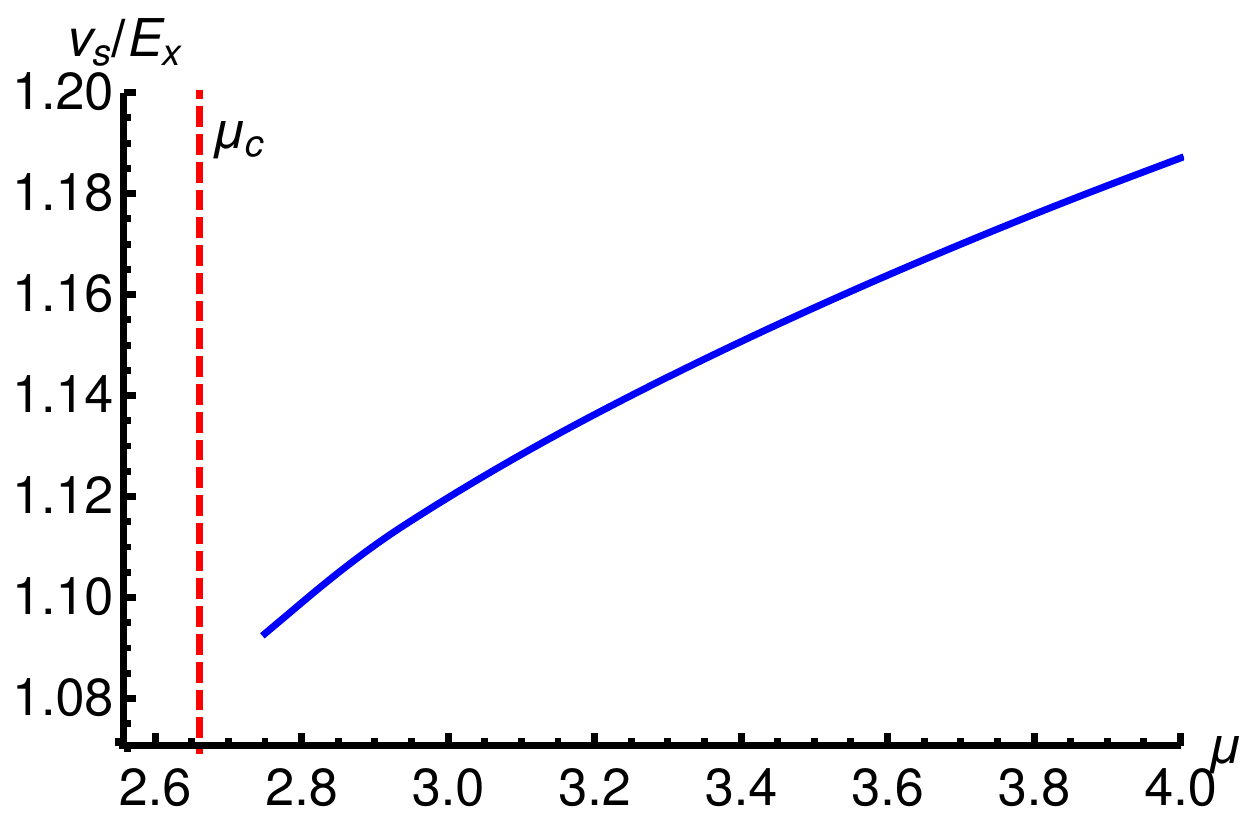}
  \caption{The dependence of the DC conductivities (left) and the velocity of the stripes $v_s$ (right) on the chemical potential. The thick solid, thick dashed, and thin solid curves are the DC conductivities perpendicular to the stripe, parallel to the stripe, and in the homogeneous solution which is subdominant for $\mu>\mu_c$, respectively. The vertical dashed red line shows the critical point $\mu=\mu_c$ in both plots.  } 
  \label{fig:DCvsmudep}
\end{figure}

\subsection{Numerical results: $\mu$-dependence}

We now proceed to study the dependence of the optical conductivities on the (rescaled) chemical potential $\mu$ and, in particular, the approach to the critical point $\mu=\mu_c \simeq 2.66$. The $\omega$-dependence of the conductivities is qualitatively similar to Fig.~\ref{fig:conductivitiesmu4} for all values of $\mu$ we have studied. Therefore, we concentrate on the $\mu$-dependence of the parameters characterizing the Drude peak. 

First we show in Fig.~\ref{fig:DCvsmudep} (left) the dependence of the DC conductivities on $\mu$. The conductivities in the striped phase (thick curves) are enhanced with respect to the subdominant homogeneous vacuum for all $\mu>\mu_c$. As the critical point is approached, all the conductivities tend to the same value, as expected because the transition is second order and the amplitude of the stripes tends to zero. We observe numerically that the differences vanish linearly as the critical point is approached from above. The same was observed for the amplitude of the CDW in~\cite{Jokela:2014dba}, whereas the amplitude of the SDW vanishes as $\sqrt{\mu-\mu_c}$. 

Apart from the Drude peak, the only structure in the longitudinal components of the optical conductivities is seen at large $\omega$, where the real parts of the conductivities approach one (see Fig.~\ref{fig:conductivitiesmu4}). We have observed that the value of $\omega$, at which the conductivities no longer differ significantly from the asymptotic value, is roughly controlled by the chemical potential, $\omega \sim \mu$.
  
In Fig.~\ref{fig:DCvsmudep} (right) we show the dependence of the velocity of the stripes $v_s$ on the chemical potential $\mu$, which turns out to be very weak. Notice that the origin of the plot is not at $v_s=0$.
 
\begin{figure}[!ht]
\center
 \includegraphics[width=0.50\textwidth]{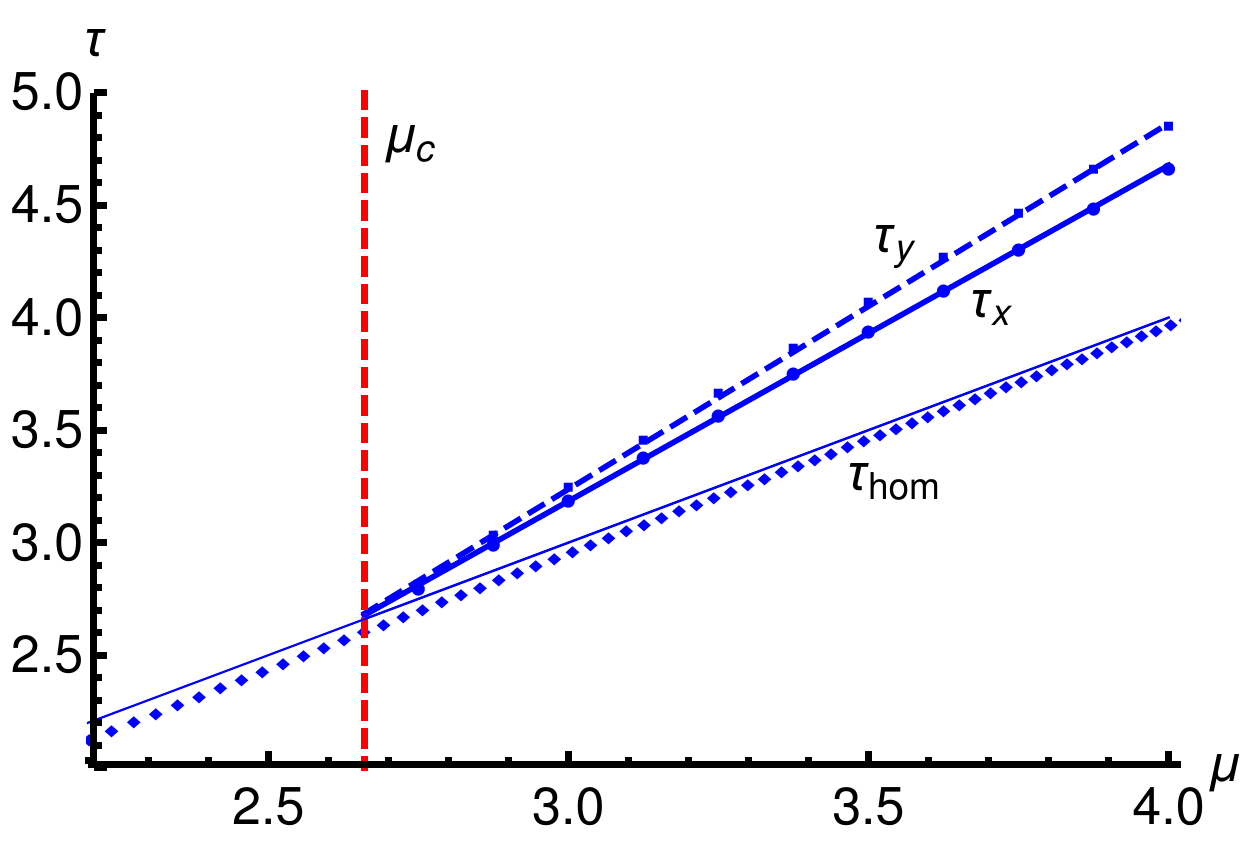}%
 \includegraphics[width=0.50\textwidth]{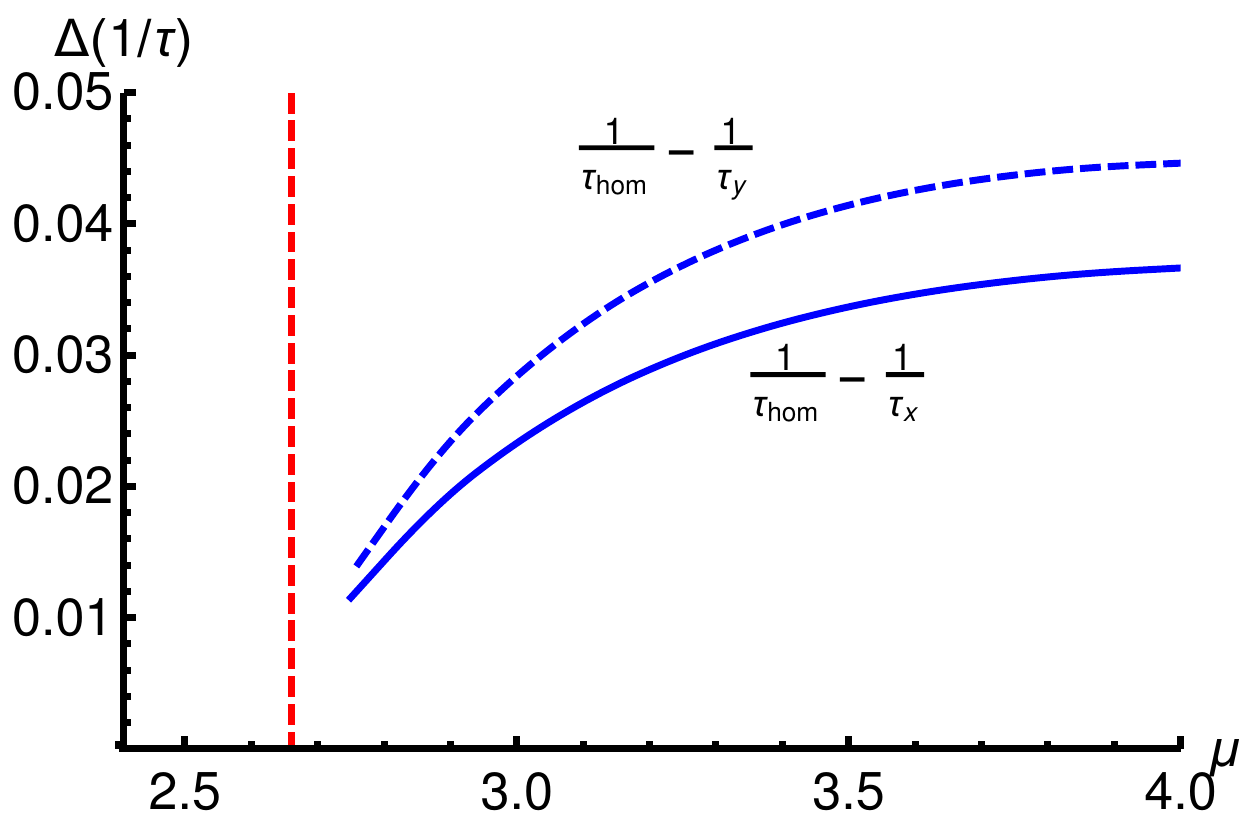}%
 \caption{The mean free time in the striped and homogeneous vacua. Left: $\tau$ as a function of $\mu$. The blue dots are our data and the solid and dashed thick lines are linear fits to $\tau(\omega)$ for the conductivity perpendicular to the stripes and parallel to the stripes, respectively. For the homogeneous vacuum we compare the mean free time $\tau_\mathrm{hom}$ to the curve $\tau=\mu$, shown as the thin line (i.e., this line is not a fit).  Right: the difference of the scattering frequencies $1/\tau$ between the homogeneous and striped solutions.} 
  \label{fig:taumudep}
\end{figure}

Finally we discuss the dependence of the mean free time $\tau$ of the Drude peaks on the chemical potentials, which is shown in Fig.~\ref{fig:taumudep} (left) for the striped and homogeneous configurations. We obtained $\tau$ as (the inverse of) the half-width of the Drude peak (rather than fitting the formula~\eqref{Drudepeak} to data). The striped vacua have longer mean free times than the homogeneous vacua, which is natural as they also had higher conductivities. The dependence of $\tau$ on $\mu$ is very close to linear for each configuration. In particular, in the homogeneous case we find that $\tau \simeq \mu$: the thin blue line in the plot shows where $\tau=\mu$ exactly. This approximation works even better at higher $\mu$  where the Drude peak becomes narrower (not shown). The thick solid and dashed blue lines show linear fits to the data for $\mathrm{Re}\,\sigma_{xx}$ and $\mathrm{Re}\,\sigma_{yy}$ in the striped phase, respectively. The fitted functions are given by
\be
 \tau_x  = 1.4917 \mu -1.2907 \ , \qquad \tau_y  = 1.6256 \mu -1.6396  \ . 
\ee

In Fig.~\ref{fig:taumudep} (right) we show the difference of the scattering frequencies between the homogeneous and striped solutions at fixed chemical potential. The plot suggests that the differences vanish nonlinearly as $\mu \to \mu_c$ from above, and the data are consistent with the behavior $\sim \sqrt{\mu-\mu_c}$, in agreement with the behavior of the amplitude of the SDW in the background.


\section{Discussion: Moving background stripes}
\label{sec:Discussion} 

An interesting question is whether a fully nonlinear solution describing moving stripes at finite velocity $v_s$ (not only infinitesimal) can be constructed. Our analysis of the DC conductivities in Sec.~\ref{sec:DC conductivity} suggest that such a solution should be sourced by a finite electric field in the $x$-direction. We restrict ourselves to the case of DC conductivity here and only discuss generic features of the solution. The numerical construction of the solution is left for future work.

In order to turn on a finite electric field, we write an Ansatz $a_x = - E_x t + \bar a_x(x-v_st,r)$ for the gauge field in $x$-direction, where $E_x$ is constant. For the other fields ($f = \psi$, $z$, $a_t$, and $a_y$) we simply write Ans\"atze of the form $f(x- v_s t,r)$. Because only derivatives of $a_x$ appear in the Lagrangian, it is immediate that explicit time dependence disappears from the equations of motion after switching to the comoving coordinate $s = x - v_s t$. We can then look for striped solutions of the system with some periodicity $L$ in $s$. A finite electric field breaks both parity and rotation symmetries, so that the solution is not expected to display any additional discrete symmetries.

Notice that in addition to the electric field, there are two parameters whose the values need to be somehow determined: the periodicity of the stripe $L$ and the velocity $v_s$. One might expect that these values are found by minimizing some  free energy of the system. We note, however, that the configuration is a nonequilibrium steady state, for which the definition of free energy is nontrivial. The finite electric field $E_x$ pumps energy into the system, resulting in a moving inhomogeneous pseudohorizon deep in the bulk.


\section{Conclusions and open problems}\label{sec:Conclusion}

In recent years, several works have used gauge/gravity duality to investigate inhomogeneous phases of finite density condensed matter systems. Phases with spontaneous striped order have particularly been in the spotlight. In view of this, it is rather surprising that the electrical conductivities have scarcely been studied. In this paper, we gave a detailed account of both the DC and AC conductivities in a top-down string theory model of dense $(2+1)$-dimensional fermionic matter with spontaneous striped order.

We studied the conductivities under the influence of small DC and AC electric fields. We were able to separate out the contributions of the modulation and the sliding to the electric conductivities. This provided us with  a new viewpoint to understand the otherwise challenging strongly coupled Fermi-like fluids which spontaneously break the translational symmetry. While most of the results that we obtained clearly conform with the physical intuition, there are some remaining puzzles. In particular, we found that the total conductivities in the direction of the applied external electric field did not, approximately, depend on the alignment with respect to stripes ($\sigma_{xx} \approx \sigma_{yy}$). 
This is a consequence of an approximate symmetry under $E_x\leftrightarrow E_y$, $\delta a_x\leftrightarrow \pm\delta a_y$, which may be linked to the SDW as we discuss in App.~\ref{app:fluct}. However, we do not understand the physical origin of this, as the stripes themselves spontaneously break the rotational symmetry.

Another outstanding issue is the mechanism underlying the persistent currents in the background geometry in the absence of any applied external electric field.  These background currents, illustrated in Fig. \ref{fig:stripy}, lead  in the spatially modulated divergence of the Hall conductivity in the $\omega \to 0$ limit, as discussed in Sec. \ref{sec:smallfreq}. One possibility is that these persistent currents are indicative of superfluidity.
However, at this point, we refrain from making such a conjecture and would like to see more evidence, say a phonon mode, to establish this interpretation. This is clearly a very interesting direction to continue our investigation of this model. One helpful generalization of the present context would be to turn on an external magnetic field, which could be relevant for relaxing a supercurrent and for demonstrating the Meissner effect. A finite background magnetic field breaks parity explicitly and would therefore also help us discern how much of the cohered charges can be attributed to the SDW component of the fluid.

We found that the sliding of the stripes is smooth and the corresponding threshold electric field for their depinning is vanishing. This is not unexpected as the gluonic sector which relaxes the momentum acts like a density of smeared impurities. 
The threshold electric field varies between different materials but, most importantly, increases with higher concentration of the impurities. We can imagine in introducing an additional optical lattice, which can be modeled holographically by a spatially dependent chemical potential. The amplitude of this additional source will play the role of the magnitude of disorder and perhaps leads to a non-vanishing threshold electric field. In the extreme case, the stripes would get strongly pinned. This sliding-pinning mechanism would provide an intriguing realization of a charge-insulator transition under very good control.
 
\addcontentsline{toc}{section}{Acknowledgments}
\paragraph{Acknowledgments}

\noindent
We would like to thank D.~Arean, O.~Bergman, A.~Donos, B.~Gout\'eraux, C.~Hoyos, E.~Keski-Vakkuri, E.~Kiritsis, R.~Meyer, G.~Policastro, M.~Rozali, A.~Yarom, and T.~Zingg for discussions.
We are especially grateful to Gilad Lifschytz for discussions and clarifications at various stages of this work.
N.~J. is supported in part by the Academy of Finland grant no. 1297472 and through the Vilho, Yrj\"o, and Kalle V\"ais\"al\"a Foundation.


\appendix

\section{Details on the analysis of DC conductivities} \label{app:DCcond}

\subsection{Comments on the gauge choice}\label{app:gauge}

It appears tempting to use the remaining gauge transformations (those left after requiring $a_u=0$) to set $p=0$ in~\eqref{Ansatz2} and~\eqref{Ansatz3}. This cannot be done, however, because we have already implicitly fixed such  gauge degrees of freedom. To illustrate this, consider a gauge transformation of the form $\delta A = d \Lambda$  with $\Lambda = q(x) t$. It would change the coefficient of the term proportional to $\log(1-u)$ in $\delta a_x$ near the boundary in the Eddington-Finkelstein coordinates, which will appear in the (standard infalling) regularity condition at the horizon. Therefore, a gauge transformation like this would imply a nontrivial change in the regularity conditions.

The gauge fixing traces back to the usual condition that $a_t$ vanishes at the horizon. Namely, as one can check, requiring the fluctuation\footnote{We are leaving out the term proportional to the derivative of the background $\partial_x a_t(x,u)$ which vanishes both at the horizon and on the boundary and plays no role here.} $\delta a_t(t,x,u) = p(x) + \delta a_t(x,u)$ to vanish at the horizon (as $\sim 1-u$) gives the same condition for $p(x)$ as the regularity of $\delta a_x$ in the Eddington-Finkelstein coordinates, which we use explicitly in Sec.~\ref{sec:DC conductivity}. 

Moreover, notice that $\delta a_t(t,x,u)$ vanishes at the horizon and $\delta a_t(x,u)$ approaches a constant on the boundary (otherwise we would be turning on a modulated electric field). Therefore, $p(x)$ is identified as the modulation of (the fluctuation of) the chemical potential, which is fully determined in terms of the electric field and the background. Only the constant part of the chemical potential can be adjusted independently of the electric field, and we set its variation to zero by requiring that the average of $p(x)$ vanishes and that $\delta a_t(x,u)$ contains no source term.

\subsection{Conserved bulk current in the presence of moving stripes}\label{app:current}

We wish to rewrite Eqs.~\eqref{curreqsvs} in a form where one immediately sees the existence of a conserved bulk current.
 We first notice that $K_1$ arises from the time derivative term of the fluctuation equation for $\delta a_x$. This term is the linear fluctuation around the background solution of the corresponding term in the generic EoM for the $a_x$ field, i.e., the fluctuation of
\be 
 -\partial_t \frac{\partial \mathcal{L}}{\partial \left(\partial_t a_x\right)} = \partial_t \frac{\partial \mathcal{L}}{\partial \left(\partial_x a_t\right)} \ ,
\ee
where we used gauge symmetry to obtain the second expression. Since, we only have linear time dependence, $K_1$ arises exactly from the terms where the time derivative operates on the combinations $v_s t$ in the fluctuation Ansatz. These terms correspond to the linearized movement of the stripes and can therefore be collected into
\be
 K_1 = - \partial_x \frac{\partial \mathcal{L}}{\partial \left(\partial_x a_t\right)}\bigg|_\mathrm{bg} = \partial_u \frac{\partial \mathcal{L}}{\partial \left(\partial_u a_t\right)}\bigg|_\mathrm{bg}\ ,
\ee
which is to be evaluated on the background solution and where the second expression follows by using the background equation of motion for $a_t$. Naturally, this result can also be checked by an explicit computation. Similarly, we find that 
\be
 K_2 =  \partial_x \frac{\partial \mathcal{L}}{\partial \left(\partial_u a_t\right)}\bigg|_\mathrm{bg} \ .
\ee
Consequently, $K_1 = \partial_u K$ and $K_2 =\partial_x K$ for
\be
K = \frac{\partial \mathcal{L}}{\partial \left(\partial_u a_t\right)}\bigg|_\mathrm{bg}
\ee
and~\eqref{curreqsvs} can be written in a form 
\be 
 \partial_u (G_1\ \partial_u \delta a_x(x,u)+v_s \tilde G_2) = 0 = \partial_x (G_1\ \partial_u \delta a_x(x,u)+v_s \tilde G_2) \ ,
\ee
where we have introduced the Routhian $\tilde G_2 = G_2 - K$.


\section{Numerical solutions and symmetries of the fluctuations}\label{app:fluct}

For completeness, we will illustrate selected plots of the numerical solutions to the fluctuation equations. From the plots one can also infer the symmetry properties discussed in the main text. We will focus on two specific cases both at nonzero electric field. 

\begin{figure}[!ht]
\center
 \includegraphics[width=0.33\textwidth]{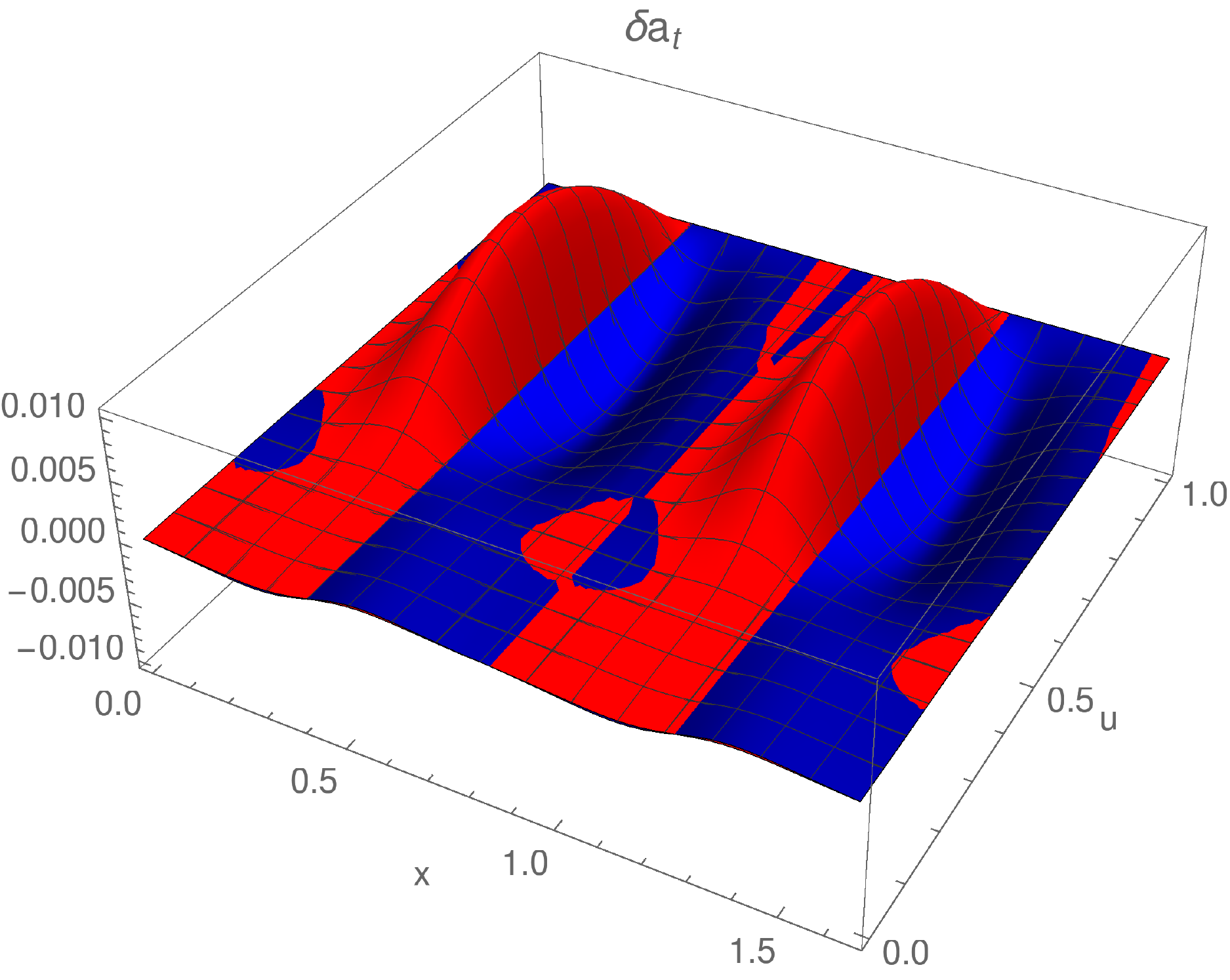}%
 \includegraphics[width=0.33\textwidth]{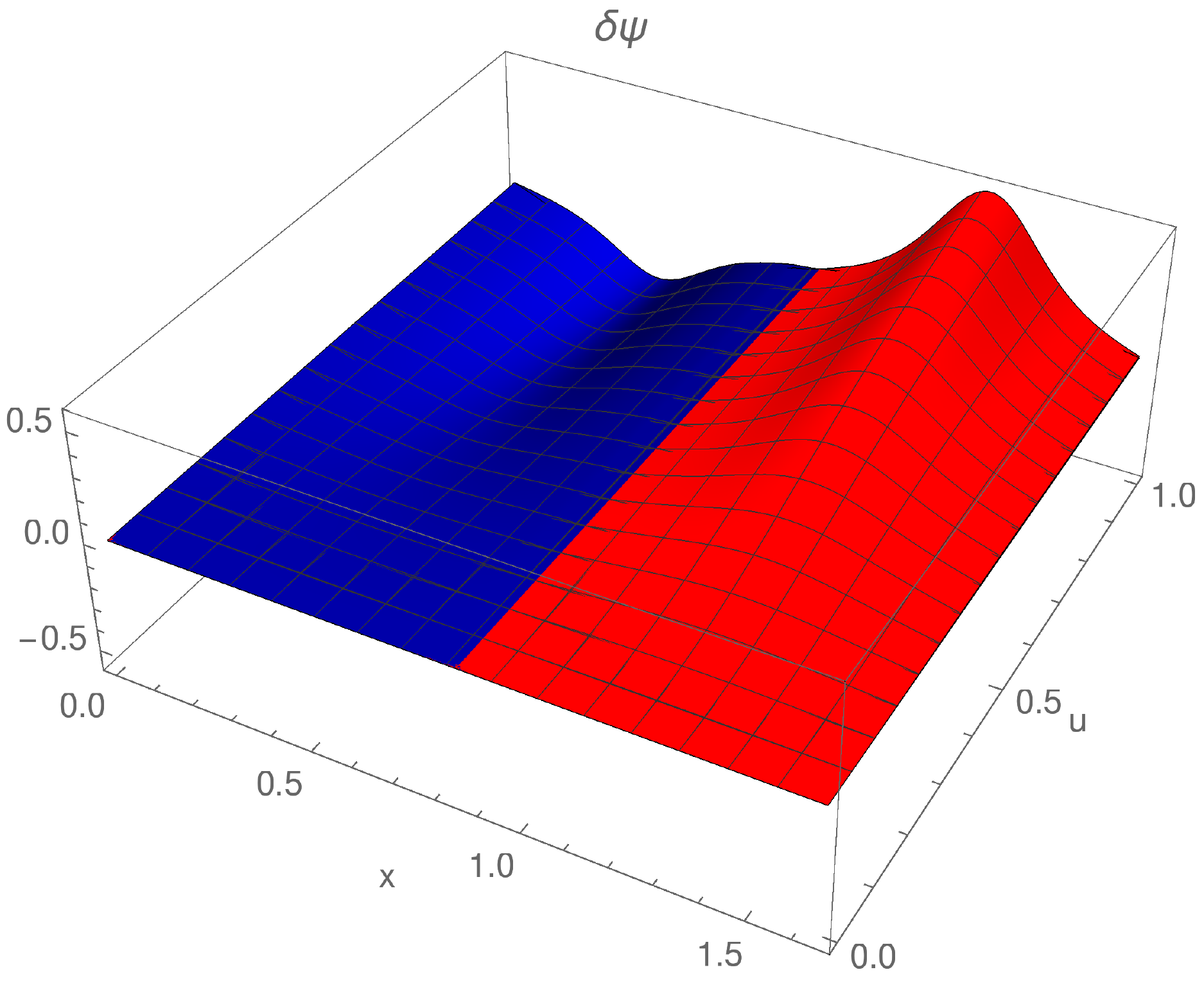}%
 \includegraphics[width=0.33\textwidth]{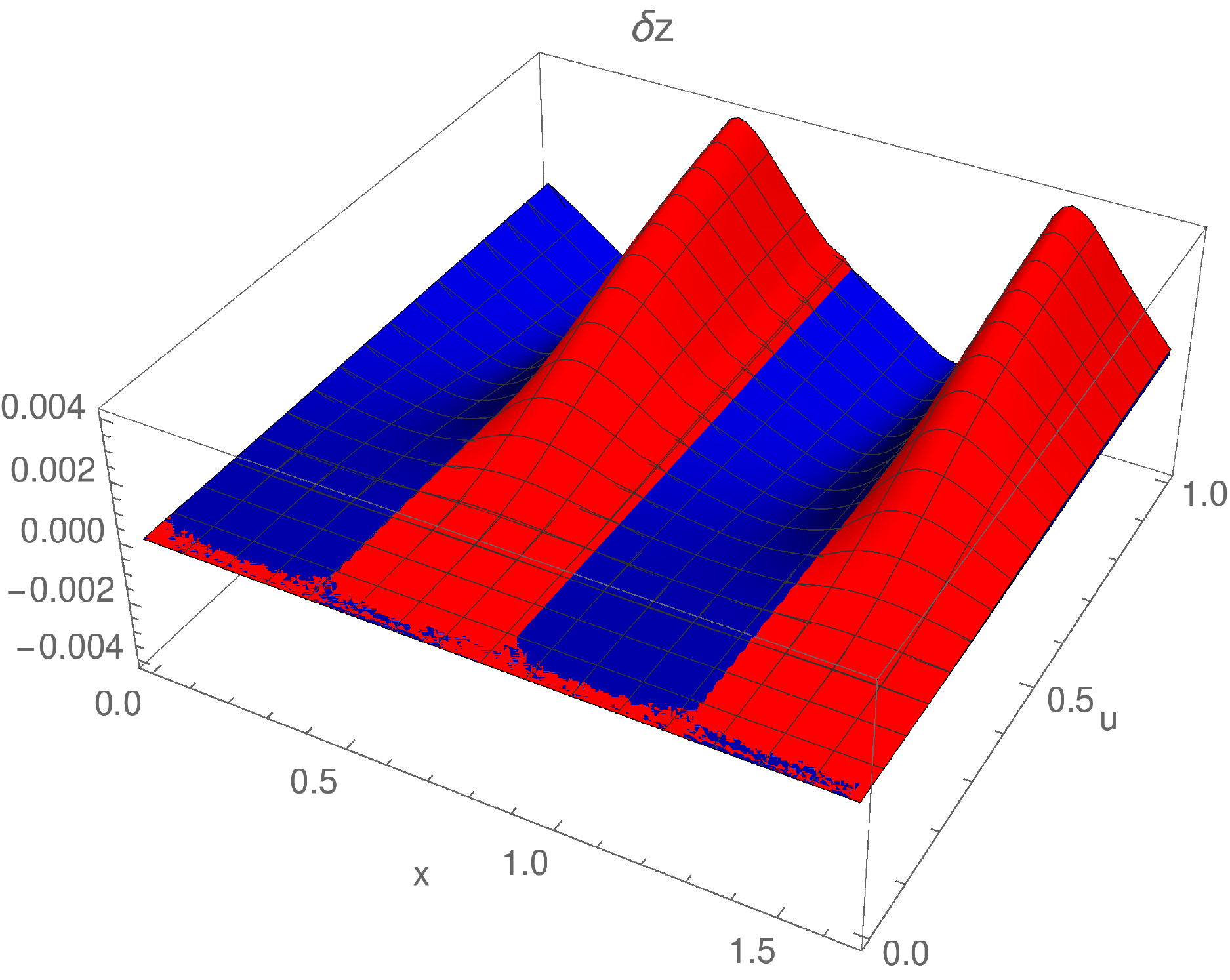}
 \includegraphics[width=0.33\textwidth]{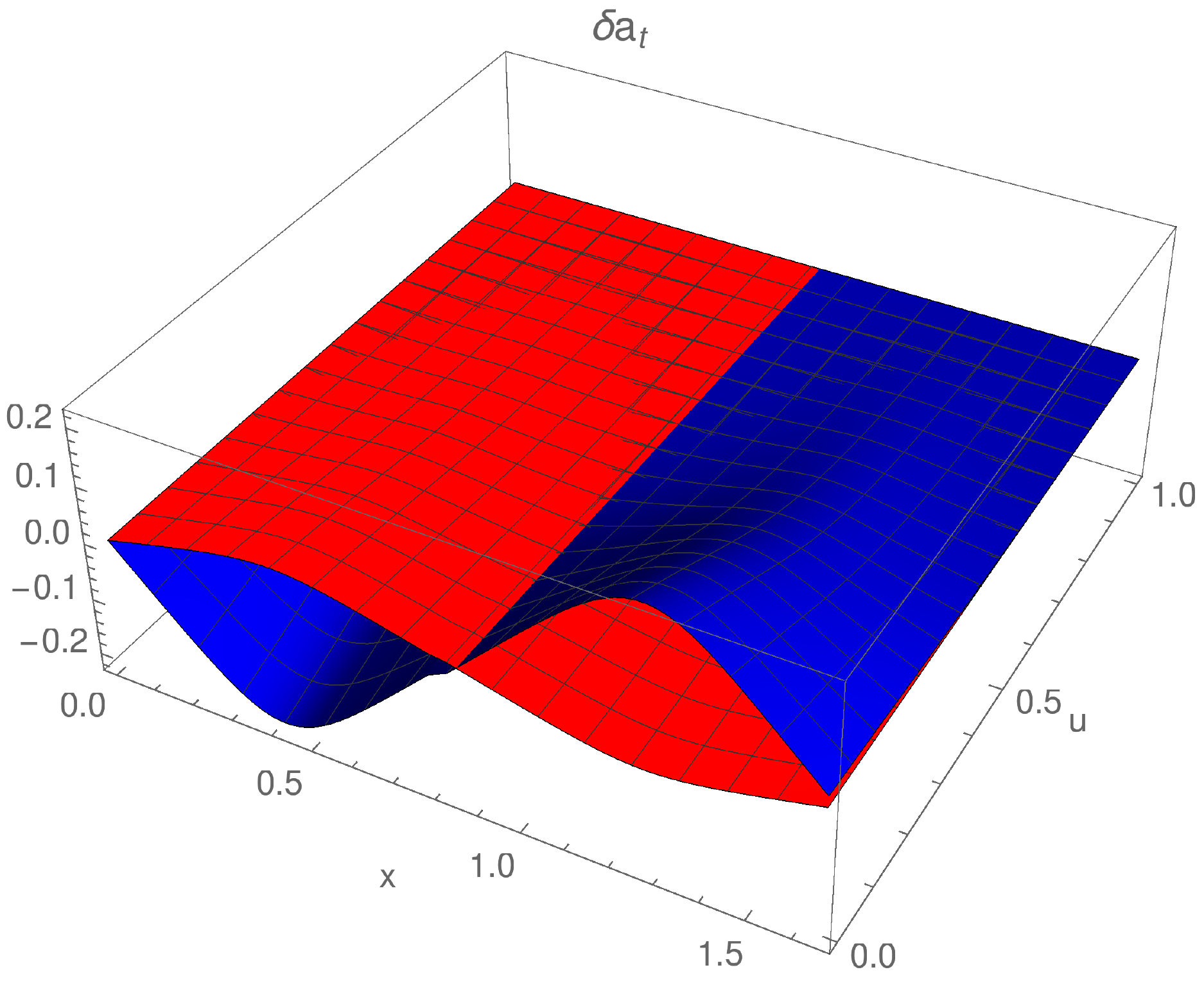}%
 \includegraphics[width=0.33\textwidth]{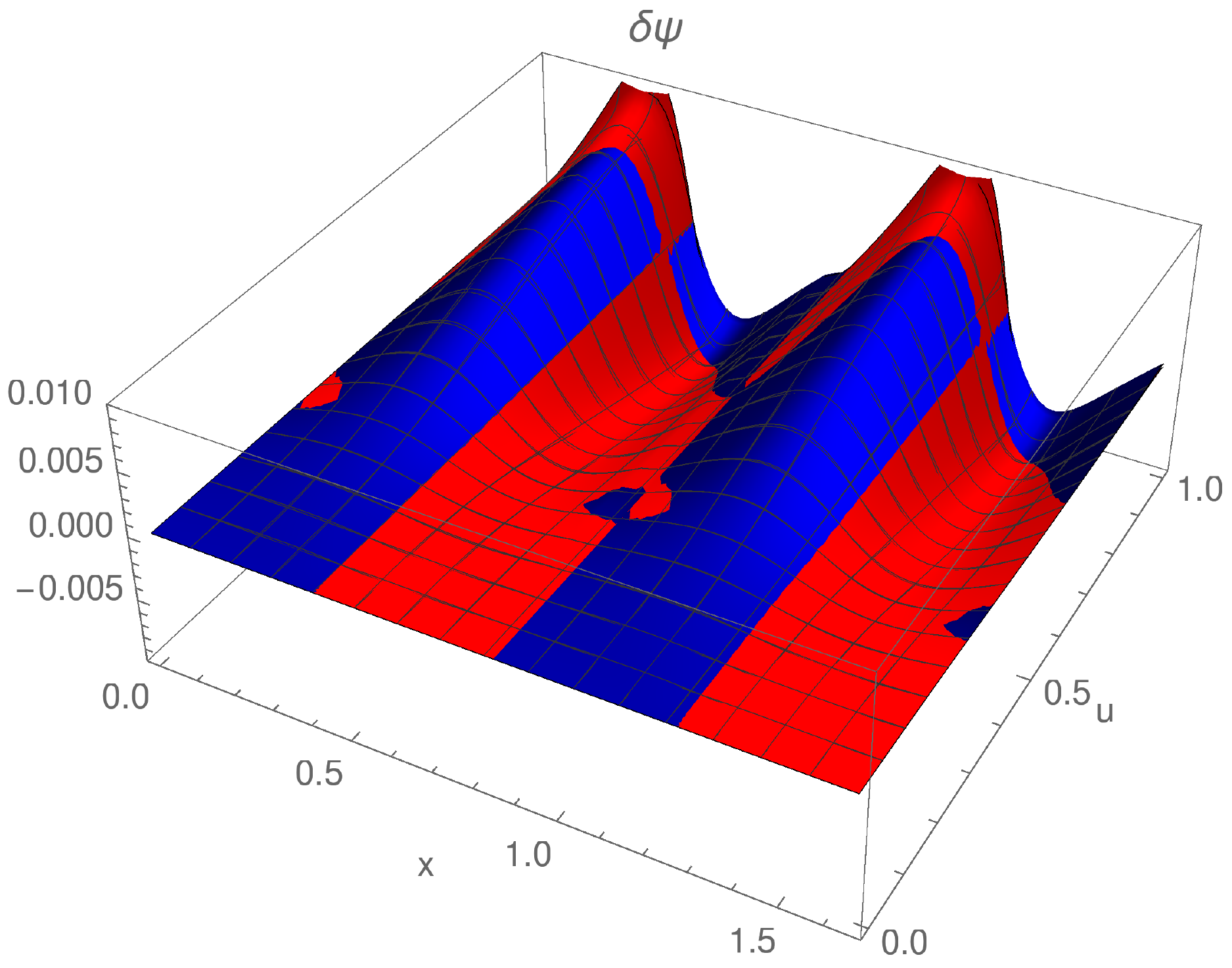}%
 \includegraphics[width=0.33\textwidth]{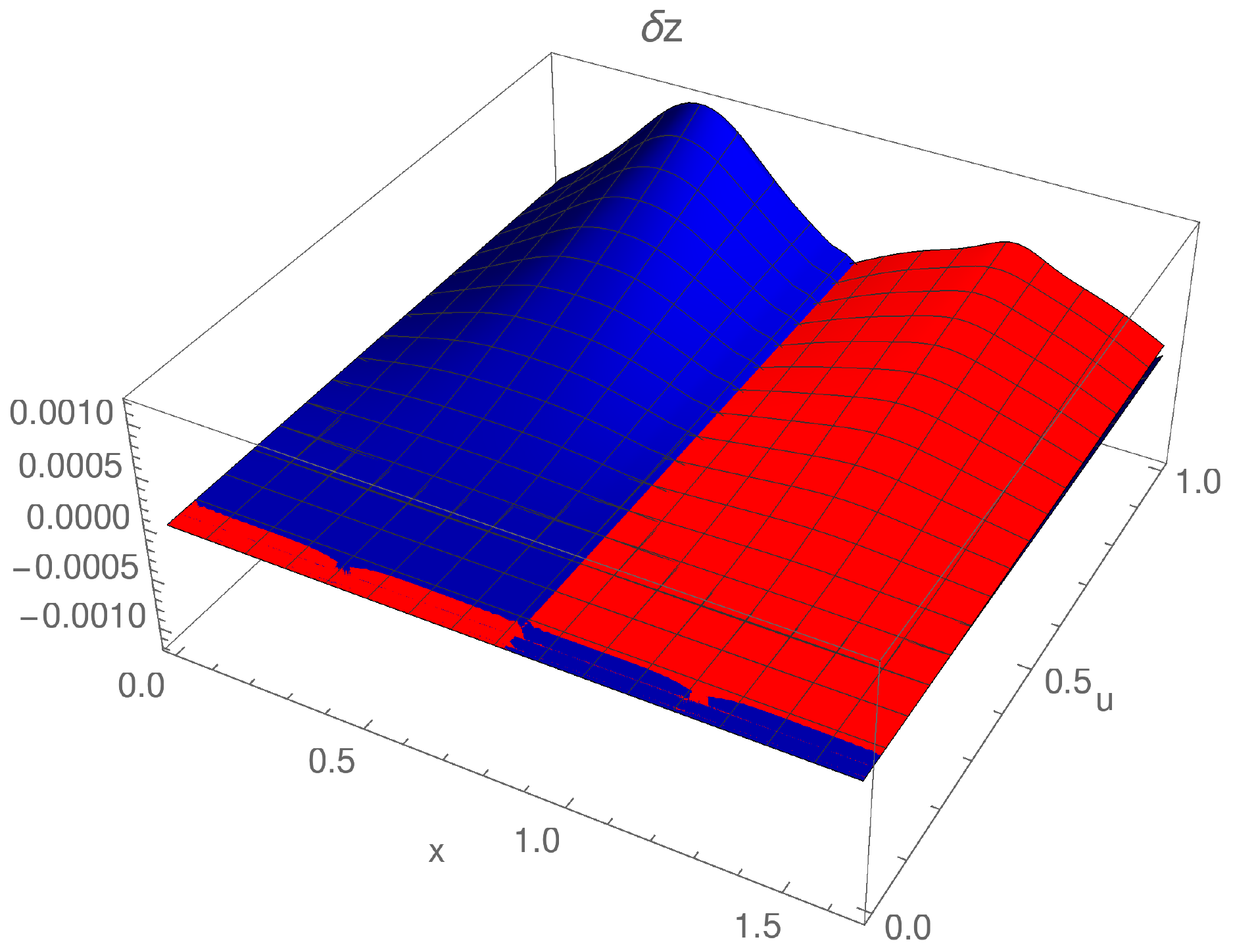}%
 \caption{The fluctuations of the temporal gauge field and the brane embedding at $\om=1$. The real (imaginary) parts of the fluctuations are given by the blue (red) surfaces. Top row: nonzero (oscillating) $E_x$. Bottom row: nonzero $E_y$.} 
  \label{fig:fluctemb}
\end{figure}
\begin{figure}[!ht]
\center
 \includegraphics[width=0.5\textwidth]{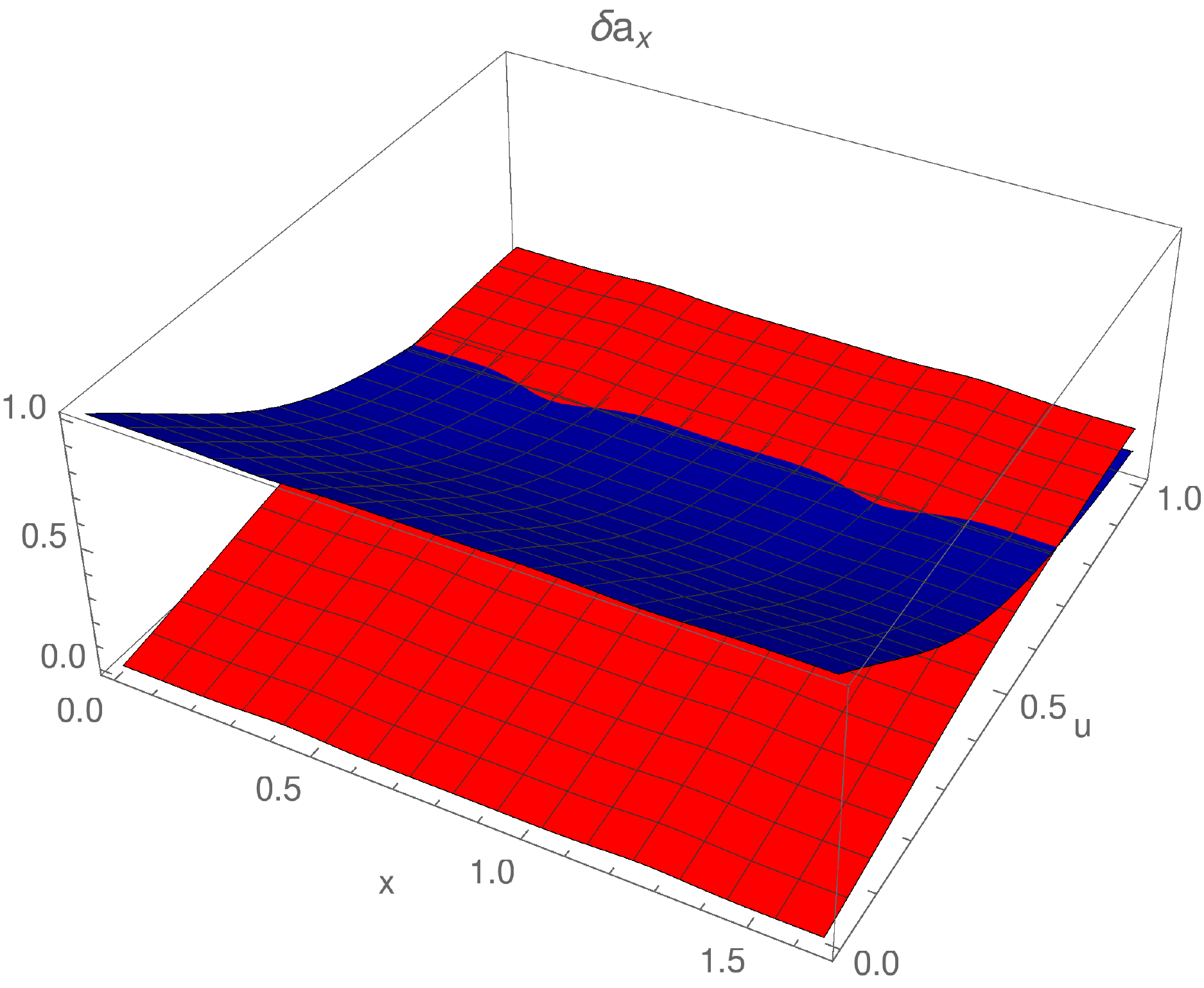}%
 \includegraphics[width=0.5\textwidth]{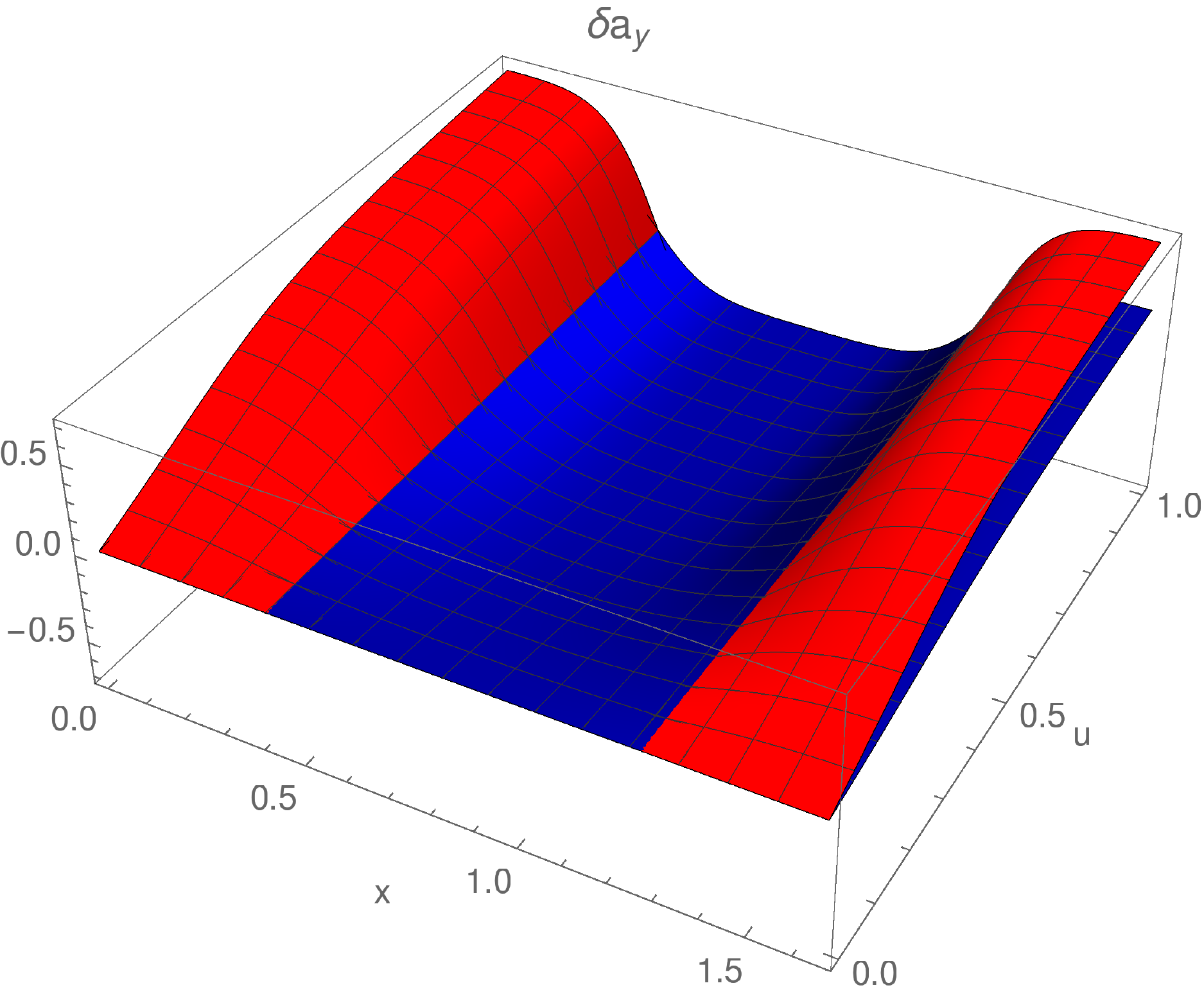}
 \includegraphics[width=0.5\textwidth]{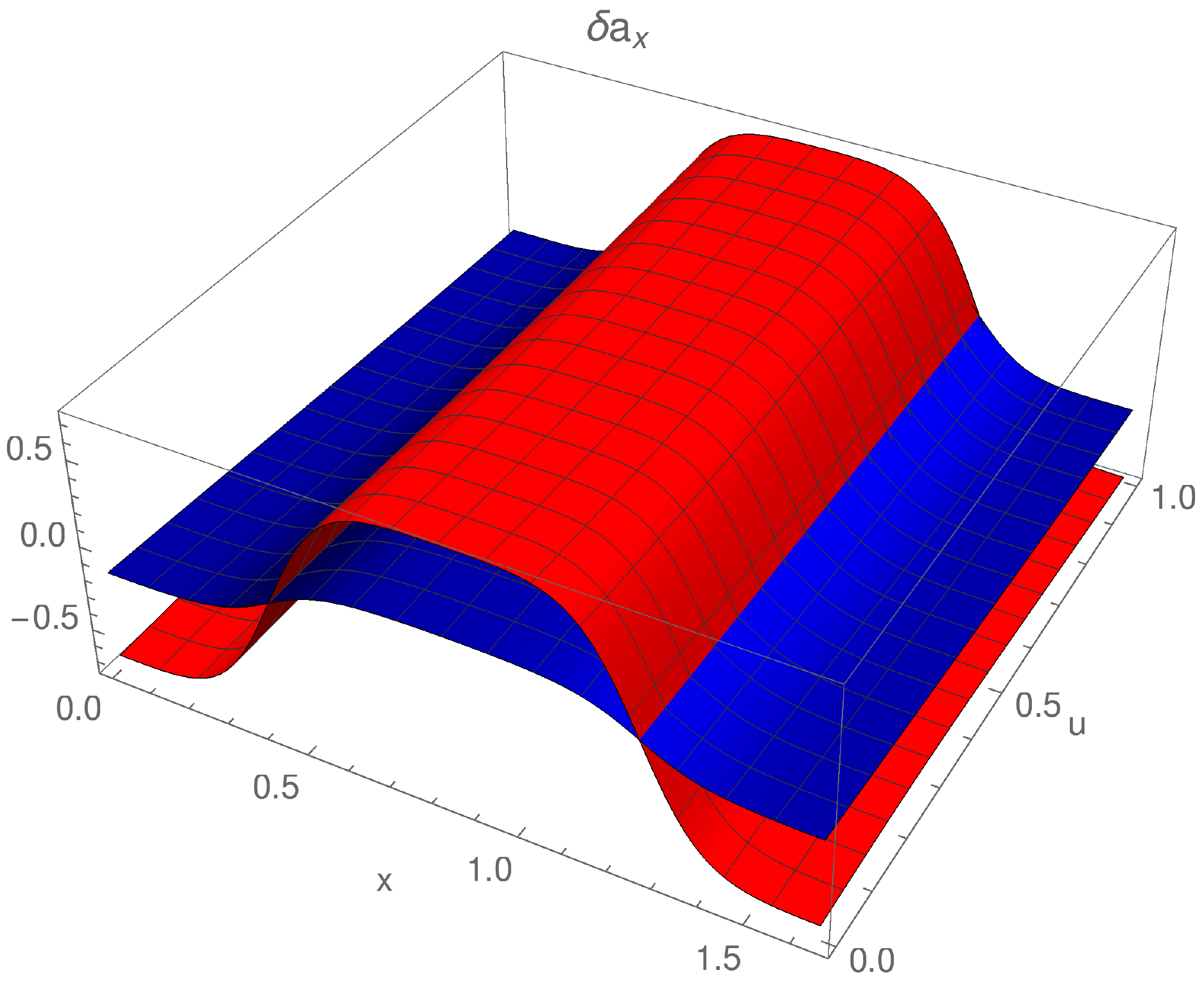}%
 \includegraphics[width=0.5\textwidth]{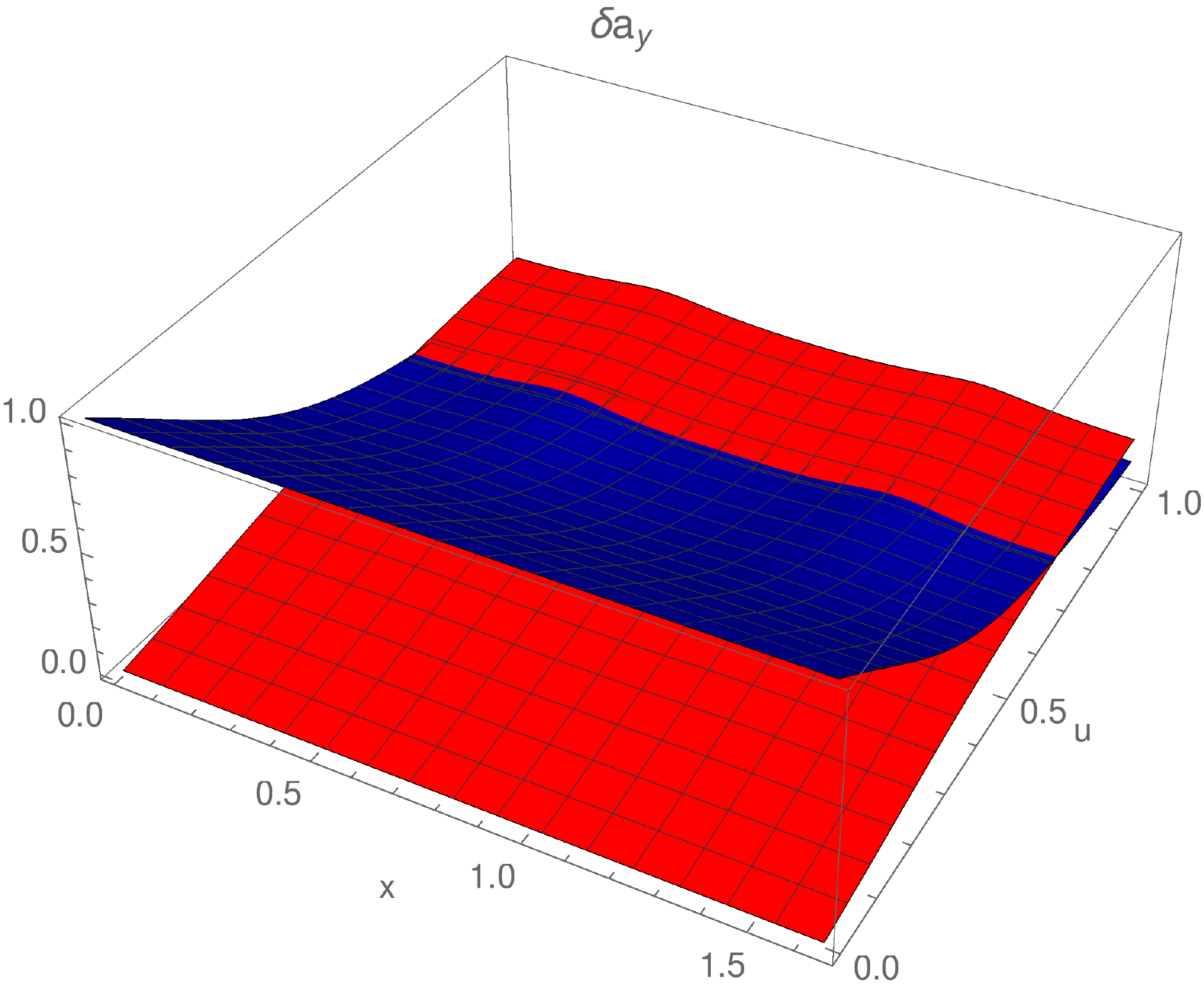}%
 \caption{The fluctuations of the gauge field components $a_x$ and $a_y$ at $\om=1$. Top row: nonzero $E_x$. Bottom row: nonzero $E_y$.} 
  \label{fig:fluctgauge_xy}
\end{figure}

We show the solutions to the fluctuation equations at frequency $\om=1$. The fluctuation of the temporal component of the gauge field $\delta a_t$ and the embedding fluctuations $\delta\psi$ and $\delta z$ are depicted in Fig.~\ref{fig:fluctemb}. For the top row plots we turn on an alternating electric field in the $x$-direction, i.e., perpendicular to the stripes, whereas in the bottom row plots are for the  case of electric field in the $y$-direction, that is along the stripes.
The gauge field fluctuations $\delta a_x$ and $\delta a_y$ are shown (with similar organization of the plots) in Fig.~\ref{fig:fluctgauge_xy}.

We observe a clear but approximate symmetry between the fluctuations for nonzero $E_x$ and $E_y$. First, both the real and imaginary part of $\delta a_x$ at nonzero $E_x$ in the top left plot of Fig.~\ref{fig:fluctgauge_xy} have similar $u$-dependence as the real and imaginary parts of $\delta a_y$ at nonzero $E_y$ in the bottom right plot. Second, $\delta a_y$ of the top right plot is very closely opposite to $\delta a_x$ in the bottom left plot, apart from the behavior near the boundary. 

Even if this symmetry is only approximate, it is surprising as the striped background solution breaks the rotation symmetry. Moreover, we notice that when the electric field is in the $x$-direction (perpendicular to the stripes), the fluctuations of $\psi$ and $a_y$ are clearly larger than those of $z$ and $a_t$, and strongly modulated. This suggests that there is a strong component of the stripe motion involved -- recall that the background is dominantly SDW, so that $\psi$ and $a_y$ are strongly modulated. When the electric field is aligned parallel to the stripes, however, the fields with large modulated fluctuations are $a_x$ and $a_t$, and the origin cannot be the motion of the stripes. Therefore, the symmetric looking configurations seemed to have essentially different origin, so that the symmetry seems quite mysterious.

A  bit more insight can be obtained by considering the $\omega \to 0$ limit. First it can be verified that the fluctuations for nonzero $E_x$ are indeed linked to the stripe motion, which can  also be seen from the leading terms in~\eqref{aydiv} and~\eqref{smallomega}. Further~\eqref{axsmallo} implies that the sizable fluctuations of $a_x$ and $a_t$ are associated with the function $p(x)$ in the limit of small $\omega$. Indeed, the coefficient of $E_y$ in the expression~\eqref{pprimegen} for $p'(x)$ turns out to be numerically much larger than the coefficients of $E_x$ and $v_s$. This is due to the strong modulation of (the background values of) $\psi$ and $a_y$ at the horizon, again linked to the SDW.

Let us finish with a heuristic analysis of the symmetry in the interchange of $x$ and $y$ for the DC conductivities, given in~\eqref{eq:sigmaxxave} and~\eqref{eq:sigmayyave}. The above discussion suggests that the symmetry is due to the SDW being dominant in our background. Therefore, we approximate that at the horizon $a_{y,0}'(x) \approx \tilde a_y\, \cos(2 \pi x/L)$ and $c(\psi_0(x)) \approx \tilde c\, \cos(2 \pi x/L)$, while $\hat \sigma$ and $a_{t,0}$ are taken to be constants. Further, motivated by the symmetry of the fluctuations in Fig.~\ref{fig:fluctgauge_xy}, we require that the corresponding coefficients of the $\omega \to 0$ limit match at the horizon in~\eqref{axsmallo} and~\eqref{aydiv}, i.e.,
\be
 \frac{p'(x)}{E_y} = \frac{v_s}{E_x} a_{y,0}'(x) \ .
\ee
Inserting here the dominant SDW contribution from~\eqref{pprimegen}, $p'(x) \approx (\sqrt{2} c/\hat\sigma - a_{t,0} a_{y,0}')E_y$, we find for the speed of the stripes $v_s/E_x \approx -a_{t,0} +\sqrt{2} \tilde c/\hat \sigma\tilde a_y$. This formula agrees with values of Fig.~\ref{fig:DCvsmudep} (right) within about 10\%. Inserting it and the above approximations in the numerically large terms in the expression~\eqref{eq:sigmaxxave} for $\langle \sigma_{xx} \rangle$ we obtain
\bea
 \langle \sigma_{xx} \rangle &\approx& \langle\hat \sigma^{-1} \rangle^{-1} +\frac{\sqrt{2}v_s}{E_x} \langle c(\psi_0) a_{y,0}'\rangle 
  -\frac{v_s}{E_x}\left\langle a_{t,0}\hat \sigma a_{y,0}'(x)^2\right\rangle \nn\\
  &\approx & \hat \sigma +\frac{1}{2 \hat\sigma} \left(\sqrt2 \tilde c - \hat\sigma a_{t,0} \tilde a_y\right)^2 \ .
\eea
This agrees with the expression for $\langle \sigma_{yy} \rangle$, after using the same approximations and dropping the numerically small\footnote{Notice that both conductivities involve terms $\propto \psi_0'^2$ which we drop here, but they also agree if we make the even rougher approximation $v_s/E_x \approx -a_{t,0} \approx 1$.}
 terms $\propto \psi_0'^2 +z_0'^2$:
\bea
 \langle \sigma_{yy} \rangle &\approx& \left\langle \hat \sigma + \frac{1}{\hat \sigma}\left(\sqrt{2}c(\psi_0) - \hat \sigma a_{t,0} a_{y,0}'\right)^2 \right\rangle \nn\\
 &\approx &   \hat \sigma +\frac{1}{2 \hat\sigma} \left(\sqrt2 \tilde c - \hat\sigma a_{t,0} \tilde a_y\right)^2 \ .
\eea
This sketch demonstrates that the approximate symmetry of the solutions is linked to the symmetry of the DC and AC conductivities observed numerically in Figs.~\ref{fig:conductivitiesmu4} and~\ref{fig:conductivitiesDClimit} and suggests that they appear because the SDW is dominant over the CDW in our background.

\end{document}